\documentclass[authoryear]{elsarticle}

\usepackage{amsmath}
\usepackage{bm}
\usepackage{tikz}
\usepackage{pgfplots}
\usepackage{color, colortbl}
\usepackage{graphicx}

\graphicspath{{figs/}}

\usepackage{hyperref}
\usepackage{multirow}
\usepackage{float}

\newproof{pf}{Proof}

\journal{arXiv.org}

\begin{document}

\begin{frontmatter}

\title{Numerical modeling of neutron transport in $\mathrm{SP_3}$ approximation by finite element method}

\author[ki]{Alexander V. Avvakumov}
\ead{Avvakumov2009@rambler.ru}

\author[nsi]{Valery~F.~Strizhov}
\ead{vfs@ibrae.ac.ru}

\author[nsi,univ]{Petr N. Vabishchevich\corref{cor}}
\ead{vabishchevich@gmail.com}

\author[univ]{Alexander O. Vasilev}
\ead{haska87@gmail.com}

\address[ki]{National Research Center \emph{Kurchatov Institute},  1, Sq. Academician Kurchatov, Moscow, Russia}
\address[nsi]{Nuclear Safety Institute, Russian Academy of Sciences, 52, B. Tulskaya, Moscow, Russia}
\address[univ]{North-Eastern Federal University, 58, Belinskogo, Yakutsk, Russia}

\cortext[cor]{Corresponding author}

\begin{abstract}
The $\mathrm{SP_3}$ approximation of the neutron transport equation allows improving the accuracy for both static and transient simulations for reactor core analysis compared with the neutron diffusion theory. 
Besides, the $\mathrm{SP_3}$ calculation costs are much less than higher order transport methods ($\mathrm{S_N}$ or $\mathrm{P_N}$). 
Another advantage of the $\mathrm{SP_3}$ approximation is a similar structure of equations that is used in the diffusion method. 
Therefore, there is no difficulty to implement the $\mathrm{SP_3}$ solution option to the multi-group neutron diffusion codes. 
In this work, the application of the $\mathrm{SP_3}$ methodology based on solution of the $\lambda$- and $\alpha$-spectral problems has been tested for the IAEA-2D and HWR reactor benchmark tests. 
The FEM is chosen to achieve the 3D geometrical generality, using GMSH as a generic mesh generator. 
The results calculated with the diffusion and $\mathrm{SP_3}$ methods are compared with the reference transport calculation results. 
It was found for the HWR reactor test that some eigenvalues are complex when calculating using both diffusion and $\mathrm{SP_3}$ options.
\end{abstract}


\end{frontmatter}

\section{Introduction}
The diffusion approximation of the neutron transport equation is widely used in nuclear reactor analysis allowing whole-core calculations with reasonable accuracy. 
The main feature of neutron diffusion equation is following: it is assumed that the neutron current is proportional to the neutron flux gradient (the Fick’s law). 
There are also three assumptions: neutron absorption much less likely than scattering, linear spatial variation of the neutron distribution and isotropic scattering \citep{stacey2007}. 
To provide the validity of diffusion theory, the modern diffusion codes use, as a rule, assembly-by-assembly coarse-mesh calculation scheme with effective homogenized cross sections, prepared by more accurate transport approximations. 
To improve the diffusion code restrictions related with limitations on mesh spacing, different approaches are used including nodal and finite element methods \citep{avvakumov2017spectral, lawrence1986progress}.

For many situations of interest (for instance, the pin-by-pin calculation taking account strongly absorbing control rods), the applicability of neutron diffusion theory are limited. 
Therefore, a more rigorous approximation for the neutron transport is required.

The solution of the neutron transport equation is very complicated problem because of seven independent variables: five for space-angular description, one for energy and one for time. 
To simplify the transport problem, different approaches are used such as the spherical harmonics ($\mathrm{P_N}$) approximation \citep{azmy2010nuclear}. 
The $\mathrm{P_N}$ approximation of the neutron transport equation is derived by expansion of the angular dependence of the neutron flux in the N spherical harmonics. 
During the last time, the simplest version of the $\mathrm{P_N}$ method, namely the simplified $\mathrm{P_N}$ approximation became widespread \citep{mcclarren2010theoretical}. 
The major feature of the $\mathrm{SP_N}$ method is following: the three-dimensional neutron transport equation is transformed to a set of one-dimensional equations. 
The number of the $\mathrm{SP_N}$ trial functions is equal to 2(N+1) compared with the $\mathrm{P_N}$ method which uses (N+1)$^2$ trial functions. 
This leads to significant reduce in the computation time for typical whole-core calculations. 

The $\mathrm{SP_N}$ approximation was first derived by Gelbard \citep{gelbard1960application, gelbard1961simplified, gelbard1962applications} in the early 1960s. 
He replaced the spatial derivatives with Laplacian and divergence operators in a one-dimensional planar geometry. 
The resulting $\mathrm{SP_N}$ equations are elliptic, for example, the $\mathrm{SP_3}$ equations consist of two equations of diffusion type with two unknown fluxes: the scalar flux and the second angular flux moment. 
More rigorous theoretical foundation of the $\mathrm{SP_3}$ methodology has been derived by Brantley and Larsen \citep{brantley2000simplified} on the basis of variational methods.

The $\mathrm{SP_3}$ method, as expected, can provide accuracy improvement compared with the common used diffusion method. 
Besides, implementation of the $\mathrm{SP_3}$ equations into the diffusion code is not difficult because of the similar structure of the $\mathrm{SP_3}$ and diffusion equations. 
For this reason the $\mathrm{SP_3}$ method was adopted in different whole-core calculation codes, such as DYN3D \citep{beckert2008development}, PARCS \citep{downar2010theory} and others. According to \citep{tada2008applicability}, application of the $\mathrm{SP_3}$ theory to the pin-by-pin calculation for BWR geometry resulted in remarkable improvement in the calculation accuracy compared with the diffusion method. 
Another report \citep{Brewster2018} shows the comparison of the diffusion and  $\mathrm{SP_3}$ methods to calculate the control rod reactivity  in a light-water reactor. As compared with the Monte Carlo reference calculation, the  $\mathrm{SP_3}$ method gives twice as accurate result compared with the diffusion method. Besides, as it turned out, the computation time using the $\mathrm{SP_3}$ method is only 1.5 times longer than that using the diffusion method \citep{tada2008applicability}.

Thus, the $\mathrm{SP_3}$ method can be considered as an improved approximation of the neutron transport equation compared with the diffusion method. 
In this regard, it will be very useful to compare the spectral parameters, calculated by both the diffusion and $\mathrm{SP_3}$ methods. 
To characterize the reactor steady-state conditions or dynamic behavior, some spectral problems are considered \citep{stacey2007, bell1970}. 
The steady-state condition is usually described by solution of a spectral problem ($\lambda$-eigenvalue problem); the fundamental eigenvalue (the largest eigenvalue) is called k-effective of the reactor core \citep{stacey2007, bell1970}. 
The reactor dynamic behavior can naturally be described on the basis of the approximate solution expansion in time-eigenvalue of $\alpha$-eigenvalue problem \citep{ginestar2002transient, verdu20103d, verdu2014modal}. 
At large times, one can talk about the asymptotic behavior of a neutron flux, whose amplitude is exp($\alpha$t). 
Because the operator matrix is nonsymmetric, some of its
eigenvalues can be complex \citep{Aragones2007FiniteDifference}.
Previously the complex eigenvalues and eigenfunctions were found in the spectral problems for some numerical tests \citep{avvakumov2017spectral}.

In this paper we consider the $\mathrm{SP_3}$ approximation for the steady-state multi-group neutron transport problem.
To solve spectral problems with nonsymmetrical matrices we use well-designed algorithms and relevant free software including the library SLEPc (Scalable Library for Eigenvalue Problem Computations, http://slepc.upv.es/). 
We use a Krylov-Schur algorithm, a variation of Arnoldi method, described in \citep{stewart2002krylov}.

The paper is organized as follows. 
The steady-state and dynamic models of a nuclear reactor based on the multigroup $\mathrm{SP_3}$ equations are given in Section 2. 
In Section 3 we discuss various spectral problems. 
Some numerical examples of calculation of spectral characteristics of two-dimensional test problems for IAEA-2D benchmark problem and HWR reactor using the two-group system of diffusion and $\mathrm{SP_3}$ equations is discussed in Section 4. 
The results of the work are summarized in Section 5.

\section{Problem statement}

Let’s consider the symmetric form of the $\mathrm{SP_3}$ equation for the neutron flux \citep{ryu2013fem}.
The neutron dynamics is considered in the limited convex two-dimensional or three-dimensional area  $\Omega$ ($\bm x = \{x_1, ..., x_d\} \in \Omega, \ d = 2,3$) with boundary $\partial \Omega$. 
The neutron transport is described by the system of equations
\begin{equation}\label{1.1}
\begin{split}
 \frac{1}{v_g} \frac{\partial \phi_{0,g}}{\partial t} - \frac{2}{v_g} \frac{\partial \phi_{2,g}}{\partial t} & -  \nabla \cdot D_{0,g} \nabla \phi_{0,g} + \Sigma_{r,g} \phi_{0,g} -  2\Sigma_{r,g} \phi_{2,g} \\ 
  & =  (1-\beta)\chi_{n,g} S_{n} + S_{s,g} + \chi_{d,g} S_d, \\
 -\frac{2}{v_g} \frac{\partial \phi_{0,g}}{\partial t} + \frac{9}{v_g} \frac{\partial \phi_{2,g}}{\partial t} & - \nabla \cdot D_{2,g} \nabla \phi_{2,g} + (5\Sigma_{t,g} + 4\Sigma_{r,g}) \phi_{2,g} - 2\Sigma_{r,g} \phi_{0,g} \\ 
  & =  -2(1-\beta)\chi_{n,g} S_{n} - 2S_{s,g} - 2\chi_{d,g} S_d,
\end{split}
\end{equation}
where
\[
S_{n} =  \sum_{g'=1}^{G} \nu \Sigma_{f,g'} \phi_{g'}, 
\quad
S_{s,g} = \sum_{g\neq g'=1}^{G} \Sigma_{s,g'\rightarrow g} \phi_{g'},
\quad
S_{d} = \sum_{m=1}^{M} \lambda_m c_m,
\]
\[
\phi_{0,g}=\phi_g + 2\phi_{2,g}, 
\quad
D_{0,g} = \cfrac{1}{3\Sigma_{tr,g}}, 
\quad
D_{2,g} = \cfrac{9}{7\Sigma_{t,g}}, 
\quad g=1,2,...,G.
\]
Here $G$ --- number of energy groups,
$\phi_g(\bm x, t)$ --- scalar flux,
$\phi_{0,g}(\bm x, t)$ --- pseudo 0th moment of angular flux,
$\phi_{2,g}(\bm x, t)$ --- second moment of angular flux,
$\Sigma_{t,g}(\bm x, t)$ --- total cross-section, 
$\Sigma_{tr,g}(\bm x, t)$ --- transport cross-section, 
$\Sigma_{r,g}(\bm x, t)$ --- removal cross-section,
$\Sigma_{s,g'\rightarrow g}(\bm x, t)$ --- scattering cross-section,
$\chi_g$  --- spectra of neutrons, 
$\nu\Sigma_{f,g}(\bm x, t)$ --- generation cross-section,
$c_m(\bm x, t)$ --- density of sources of delayed neutrons,
$\lambda_m$ --- decay constant of sources of delayed neutrons,
$M$ --- number of types of delayed neutrons.

The density of sources of delayed neutrons is described by the equations
\begin{equation}\label{1.2}
 \frac{\partial c_m}{\partial t} + \lambda_m c_m = \beta_m S_{n},
 \quad m = 1,2, ..., M, 
\end{equation}
where $\beta_m$ is the fraction of delayed neutrons of m-type, and
\[
 \beta = \sum_{m=1}^{M} \beta_m.
\] 
The Marshak-type conditions are set at the boundary of the area $\partial \Omega$:
\begin{equation}\label{1.3}
\begin{split}
\begin{bmatrix}
J_{0,g}(\bm x)\\
J_{2,g}(\bm x)\\
\end{bmatrix}
& =
\begin{bmatrix}
\phantom{-}\cfrac{1}{2} & -\cfrac{3}{8} \\
 -\cfrac{3}{8} & \phantom{-}\cfrac{21}{8} \\
\end{bmatrix}
\begin{bmatrix}
\phi_{0,g}(\bm x) \\
\phi_{2,g}(\bm x) \\
\end{bmatrix},
\\
J_{i,g}(\bm x) & = -D_{i,g}\nabla\phi_{i,g}(\bm x), 
\quad
i = 0, 2.
\end{split}
\end{equation}

System of equations (\ref{1.1}) and (\ref{1.2}) is supplemented with boundary conditions (\ref{1.3}) and corresponding initial conditions:
\begin{equation}\label{1.4}
 \phi_g(\bm x,0) = \phi_g^0(\bm x), 
 \quad g = 1,2, ..., G,
 \quad c_m(\bm x,0) = c_m^0(\bm x), 
 \quad m = 1,2, ..., M.
\end{equation}

Let's write the boundary problem (\ref{1.1})--(\ref{1.4}) in operator form. 
The vectors $\bm u_1 = \{\phi_{0,1}, \phi_{0,2}, \cdots, \phi_{0,G}\}$, $\bm u_2 = \{\phi_{2,1}, \phi_{2,2}, \cdots, \phi_{2,G}\}$, $\bm c = \{c_1, c_2, ..., c_M\}$ and matrices are defined as follows
\[
\begin{split}
V & = (v_{gg'}),
\quad 
v_{gg'} = \frac{1}{v_g} \delta_{gg'}, \\
B & = (b_{gg'}),
\quad 
b_{gg} = -2\Sigma_{r,g},
\quad 
b_{gg'} = 2\Sigma_{s, g' \rightarrow g}, \\
A_1 & = (a_{gg'}),
\quad 
a_{gg} = -\nabla \cdot D_{0,g} \nabla + \Sigma_{r,g},
\quad 
a_{gg'} = -\Sigma_{s, g' \rightarrow g}, \\
A_2 & = (a_{gg'}),
\quad 
a_{gg} = -\nabla \cdot D_{2,g} \nabla + 5\Sigma_{tr,g} + 4\Sigma_{r,g},
\quad 
a_{gg'} = -4\Sigma_{s, g' \rightarrow g}, \\
F & = (f_{gg'}),
\quad 
f_{gg'} = \chi_{n,g}\nu\Sigma_{f,g'}, \\
E & = (e_{gm}),
\quad 
e_{gm} = \chi_{d,g}\lambda_m, \\
\Lambda & = (\lambda_{mm'}), 
\quad 
\lambda_{mm'} = \delta_{mm'}\lambda_m, \\
Q & = (q_{mg}),
\enskip
q_{mg} =\beta_m \nu\Sigma_{f,g},
\end{split}
\]
where
\[
 \delta_{g g'} = \left \{ 
 \begin{matrix}
 1, & g = g', \\
 0, & g \neq  g',
 \end{matrix}
 \right. 
\]  
is the Kronecker symbol.
We shall use the set of vectors $\bm u$, whose components satisfy the boundary conditions (\ref{1.3}). 
Using the set definitions, the system of equations (\ref{1.1}) and (\ref{1.2}) can be written as following
\begin{equation}\label{1.5}
\begin{split}
V \left (\frac{d \bm u_1}{d t} - 2 \frac{d \bm u_2}{d t} \right ) + A_1 \bm u_1 + B \bm u_2 &=(1-\beta) F (\bm u_1 - 2\bm u_2) + E\bm c,
\\
V \left (- 2 \frac{d \bm u_1}{d t} + 9 \frac{d \bm u_2}{d t} \right )  + B \bm u_1 + A_2 \bm u_2 &=-2(1-\beta) F (\bm u_1 - 2\bm u_2) - 2E\bm c,
\\
\frac{d \bm c}{d t} + \Lambda \bm c &= Q (\bm u_1 - 2\bm u_2). 
\end{split}
\end{equation}
Without taking into account delayed neutrons (all neutrons are considered as prompt), we have
\begin{equation}\label{1.6}
\begin{split}
V \left (\frac{d \bm u_1}{d t} - 2 \frac{d \bm u_2}{d t} \right ) + A_1 \bm u_1 + B \bm u_2 &= F (\bm u_1 - 2\bm u_2),
\\
V \left ( - 2 \frac{d \bm u_1}{d t} + 9 \frac{d \bm u_2}{d t} \right ) + B \bm u_1 + A_2 \bm u_2 &=-2 F (\bm u_1 - 2\bm u_2).
\end{split}
\end{equation}
The Cauchy problem is formulated for equations  (\ref{1.6}) when
\begin{equation}\label{1.7}
 \bm u_1(0) = \bm u_1^0, \quad  \bm u_2(0) = \bm u_2^0, 
\end{equation} 
where $\bm u_1^0 = \{\phi_{0,1}^0,  \phi_{0,2}^0, ...,  \phi_{0,G}^0 \}$, 
$\bm u_2^0 = \{\phi_{2,1}^0,  \phi_{2,2}^0, ...,  \phi_{2,G}^0 \}$. 
Для уравнений (\ref{1.5})  задается также начальное условие
\begin{equation}\label{1.71}
 \bm c(0) = \bm c^0,
\end{equation} 
where $\bm c^0 = \{ c_1^0,  c_2^0, ...,  c_M^0 \}$.

\section{Spectral problems}

To characterize the reactor dynamic processes described by
Cauchy problem (\ref{1.5})-(\ref{1.7}), let’s consider some spectral problems \citep{bell1970,stacey2007}.

The spectral problem, which is known as the $\lambda$-spectral problem, is usually considered.
For the system of equations (\ref{1.6}), (\ref{1.7}), we have
\begin{equation}\label{1.8}
L \bm \varphi = \lambda^{(k)} M \bm \varphi,
\end{equation}
where
\[
\bm \varphi = \{\bm \varphi_1, \bm \varphi_2\},
\quad
L = \begin{pmatrix}
A_1 & B \\
B & A_2 \\
\end{pmatrix},
\quad
M = \begin{pmatrix}
F & -2F \\
-2F & 4F \\
\end{pmatrix}.
\]
The minimal eigenvalue is used for characterisation of neutron field, thus
\[
 k = \frac{1}{\lambda^{(k)}_1}  
\] 
is the effective multiplication factor (k-effective).
The value $k = \lambda^{(k)}_1 = 1$ is related to the critical state of the reactor, and the corresponding eigenfunction $\bm{\varphi}^{(1)}(\bm x)$ is the stationary solution of the Eq (\ref{1.5}), (\ref{1.6}).
At $k > 1$, one can speak about supercriticality, at $k < 1$ --- about subcriticality.

The spectral problem (\ref{1.8}) cannot directly be connected with the
dynamic processes in a nuclear reactor. The eigenvalues of the multiplication factor of the reactor and the corresponding eigenfunctions do not depend on the time delay for the emission of delayed neutrons. 
The reason is that the problem (\ref{1.8}) on eigenvalues is the problem of finding time-independent solutions of the neutron transport equation, and the term describing the contribution of fission to the neutron balance is equal to the total number of fission neutrons, both instantaneous and delayed divided by $k$.
At the best, we can get only the limiting case --- the stationary critical state.
The more acceptable spectral characteristics for the non-stationary equation (\ref{1.5}) are related the spectral problem
\begin{equation}\label{1.9}
\begin{split}
L \bm \varphi - (1 - \beta) M \bm \varphi - I \bm s &= \lambda^{(\alpha)} W \bm \varphi, \\
\Lambda \bm s - R \bm \varphi  &= \lambda^{(\alpha)} \bm s.
\end{split}
\end{equation}
The  $\alpha$-eigenvalue problem without delayed neutrons (\ref{1.6}) can be written as follows
\begin{equation}\label{1.10}
L \bm \varphi - M \bm \varphi = \lambda^{(\alpha)} W \bm \varphi,
\end{equation}
where
\[
I = \begin{pmatrix}
E \\
-2E \\
\end{pmatrix},
\quad
R = \begin{pmatrix}
Q & -2Q \\
\end{pmatrix},
\quad
W = \begin{pmatrix}
V & -2V \\
-2V & 9V \\
\end{pmatrix}
\]
The fundamental eigenvalue
\[
 \alpha = \lambda^{(\alpha)}_1
\]
is called \citep{bell1970} the $\alpha$-eigenvalue or the period eigenvalue, because it is inversely related to the reactor period.
The problem of the period eigenvalues essentially takes into account the contribution of delayed neutrons.
In particular, the long lifetime of the predecessors of delayed neutrons makes a large contribution to the slowly decreasing eigenfunctions of the reactor period, and this does not occur when only instantaneous neutrons are taken into account.

The asymptotic behaviour of Cauchy problem solution (\ref{1.5})-(\ref{1.7}) at large times can be connected with the eigenvalue $\alpha$.
In this regular mode, the reactor behaviour is described by the function $\exp(-\alpha t) \bm \varphi^{(1)}(\bm x)$.
If $\alpha = 0$, then the reactor is critical; if $\alpha > 0$, then we get the neutron flux decreasing (subcritical state), and if $\alpha <  0$, then we get the neutron flux increasing (supercritical state).

There is a simple approximate relationship between the dominant eigenvalues of the $k$-spectral ($\lambda$-spectral) and $\alpha$-spectral problems without delayed neutrons \citep{verdu20103d}:
\begin{equation}\label{1.11}
\ \dfrac{1 - k}{k} \approx \mathop{\alpha} \Lambda_{pr},  
\ 1 - k \approx \mathop{\alpha} \l_{pr},
\end{equation}
where $\Lambda_{pr}$ ---  the prompt neutron generation time and  $l_{pr}$ ---  the prompt neutron lifetime. The prompt neutron generation time is determined as the neutron flux functional using adjoint fundamental $\alpha$-eigenfunction rather then the $ k $-eigenfunction \citep{verdu20103d}.

For the spectral problems with delayed neutrons, the relationship is as folows
\begin{equation}\label{1.12}
\ \dfrac{1 - k}{k} \approx \mathop{\alpha_{tot}} \Lambda_{pr} + \mathop{\alpha_{tot}} \sum_{m=1}^{M} \dfrac{\beta_m}{\lambda_m - \alpha_{tot}},
\end{equation}
which corresponds to inhour's equation. 
It is obvious that if we consider small perturbations of reactivity ($ | k - 1 | < \beta$), then the dominant $\alpha_{tot}$-eigenvalue does not practically depend on $\Lambda_{pr}$ and is related  only with delayed neutrons parameters $ \lambda_m $ and $ \beta_m $. Under such assumptions one can expect that the shapes of the fundamental eigenfunctions of different spectral problems are similar and, therefore, the relationships (\ref{1.11}) and (\ref{1.12}) are close to exact solution. In this case, we can derive the following relationship between $\alpha$ and $\alpha_{tot}$.
\begin{equation}\label{1.13}
\ \alpha \approx  \dfrac{\alpha_{tot}}{\Lambda_{pr}} \sum_{m=1}^{M} \dfrac{\beta_m}{\lambda_m - \alpha_{tot}} + \alpha_{tot},
\end{equation} 
where the second term in the rihght part of the equation (\ref{1.13}) can be neglected.

It is supposed that the relationships (\ref{1.11})-(\ref{1.13}) are valid for the rest eigenvalues of the $k$-spectral and $\alpha$-spectral problems \citep{verdu20103d}.

\section{Numerical examples}

To study the properties of the eigenvalues and eigenfunctions of dufferent types, several benchmarks are studied.
The first benchmark is the IAEA-2D two dimensional hexagonal problem of VVER-type without reflector \citep{chao1995}.
The second benchmark is the similar test but with radial reflector \citep{chao1995}. 
The aim of the third test is to investigate azimutally non-symmetric effects on the eigenvalues and eigenfunctions 
(the changed IAEA-2D test with reflector).
Finally, we studied the complex eigenvalues and eigenfunctions for the heavy water hexagonal reactor test HWR \citep{chao1995}.

\begin{figure}[h]
	\begin{minipage}{0.30\linewidth}
		\center{\includegraphics[width=1\linewidth]{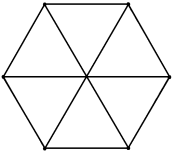}}\\
	\end{minipage}
	\hfill
	\begin{minipage}{0.30\linewidth}
		\center{\includegraphics[width=1\linewidth]{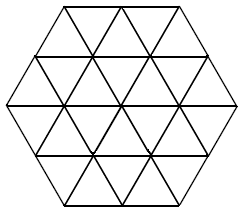}}\\
	\end{minipage}
	\hfill
	\begin{minipage}{0.30\linewidth}
		\center{\includegraphics[width=1\linewidth]{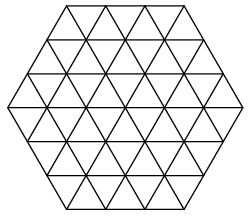}}\\
	\end{minipage}
	\caption{Discretization of assembly into 6, 24 and 96 finite elements.}
	\label{fig:mesh}
\end{figure}

The two-group model ($G = 2$) is used in all tests. 
The method of finite elements \citep{brenner2008, quarteroni2008} on triangular calculation grids is used for the approximate solution of the spectral problem. 
The standard Lagrangian finite elements are used.
The software has been developed using the engineering and scientific calculation library FEniCS \citep{logg2012}.
SLEPc has been used for numerical solution of the spectral problems.
We used a Krylov-Schur algorithm with an accuracy of $10^{-15}$.
The following parameters were varied in the calculations:
\begin{itemize}\itemsep1pt \parskip0pt \parsep0pt
\item $n$ --- the number of triangles per one assembly (Fig.~\ref{fig:mesh}); 
\item $p$ --- the order of finite element.
\end{itemize}
In this work we compare the $\mathrm{SP_3}$ calculations with the previous diffusion model calculations \citep{avvakumov2014, avvakumov2017spectral}.

\subsection{IAEA-2D without reflector}

\begin{figure}[h]
	\center{\includegraphics[width=0.7\linewidth]{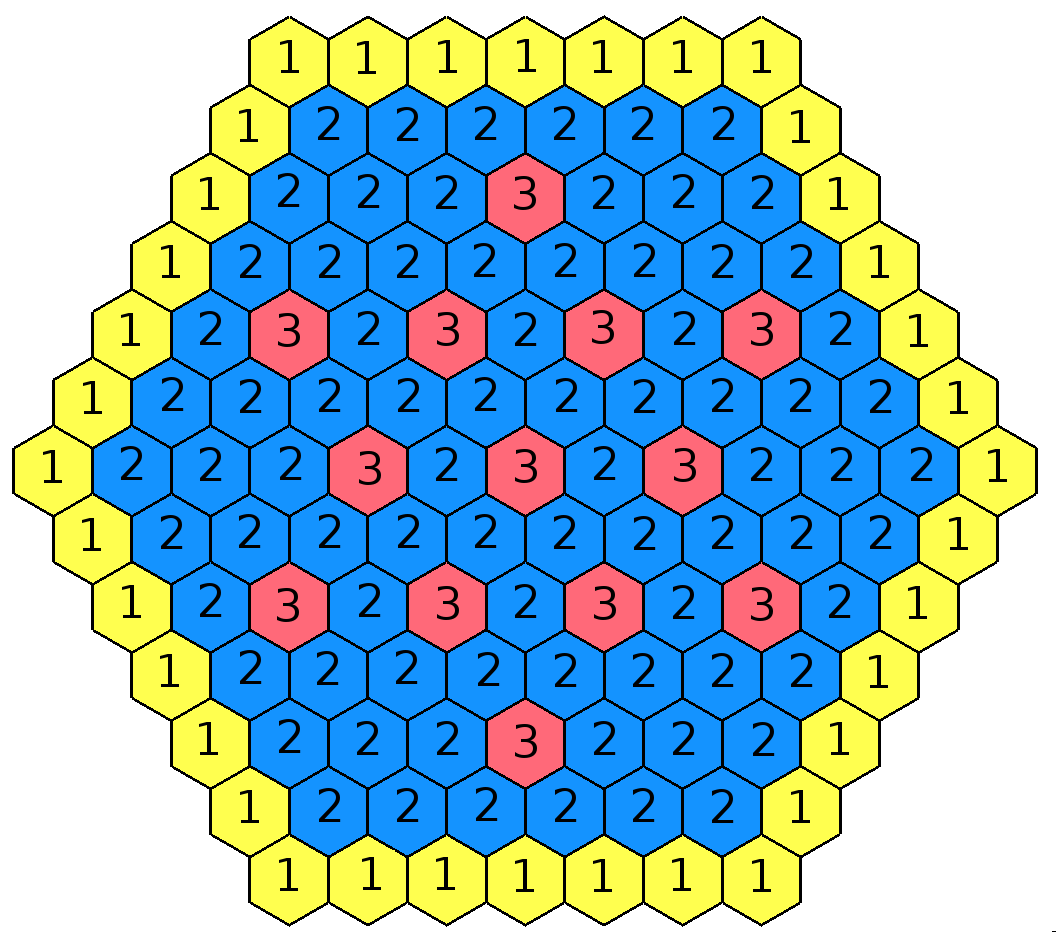}}\\
	\caption{Geometrcial model of the IAEA-2D reactor core without reflector.}
	\label{fig:iaea}
\end{figure}

The geometrical model of the IAEA-2D reactor core \citep{chao1995} consists of a set of hexagonal assemblies and is presented in Fig.~\ref{fig:iaea}, where the assemblies of various types are marked with various digits. 
The total size of assembly equals 20 cm. 

Diffusion neutronics constants in the common units are given in Table~\ref{tab:iaea}.
The following delayed neutrons parameters are used: one group of delayed neutrons with effective fraction $\beta_1 = 6.5\cdot10^{-3}$ and decay constant $\lambda_1 = 0.08$ s$^{-1}$. 
Neutron velocity  $v_1 = 1.25 \cdot 10^7$ cm/s and $v_2 = 2.5 \cdot 10^5$ cm/s.

\begin{table}[h]
\caption{Diffusion neutronics constants for IAEA-2D.}
\label{tab:iaea}
\begin{center}
\begin{tabular}{c l l l l l}
\hline
Material & 1 & 2 & 3 & 4\\
\hline 
$D_1$ & 1.5 & 1.5 & 1.5 & 1.5\\
$D_2$ & 0.4 & 0.4 & 0.4 & 0.4\\
$\Sigma_{a1}$ & 0.01 & 0.01 & 0.01 & 0.0\\
$\Sigma_{a2}$ & 0.08 & 0.085 & 0.13 & 0.01\\
$\Sigma_{s,1\rightarrow2}$ & 0.02 & 0.02 & 0.02 & 0.04\\
$\Sigma_{s,1\rightarrow1}$ & 0.1922222 & 0.1922222 & 0.1922222 & 0.1822222\\
$\Sigma_{s,2\rightarrow2}$ & 0.7533333 & 0.7483333 & 0.7033333 & 0.8233333\\
$\nu_1\Sigma_{f1}$ & 0.00 & 0.00 & 0.00 & 0.00\\
$\nu_2\Sigma_{f2}$ & 0.135 & 0.135 & 0.135 & 0.00\\
\hline
\end{tabular}
\end{center}
\end{table}

\subsubsection{Solution of Lambda Modes spectral problem}

As a reference solution for the diffusion model, we used the previous results \citep{avvakumov2014}; for the $\mathrm{SP_3}$ model --- the solution obtained using very fine mesh ($p = 3, \  n = 96$). The maximum difference in assembly power between two models is about 2 percent for the rodded assemblies (material 3, see Fig.~\ref{fig:iaea}).

The results of the solution of the effective multiplication factor for test IAEA-2D without a reflector are shown in Table~\ref{tab:iaea_without_lambda}.
Hereinafter, for $\lambda$-spectral problems, the following notation is used: $k_{dif}$ --- effective multiplication factor for the diffusion model; $k_{sp_3}$ --- effective multiplication factor for the $\mathrm{SP_3}$ model; $\Delta$ --- absolute deviation from the reference value in pcm ($10^{-5}$); $\delta$ --- the standard deviation of the relative power in percent. 
These data demonstrate the convergence of the computed eigenvalues with refinement of the calculation mesh and increase in polynomial degree.

\begin{figure}[H]
\begin{center}
	\includegraphics[width=0.75\linewidth]{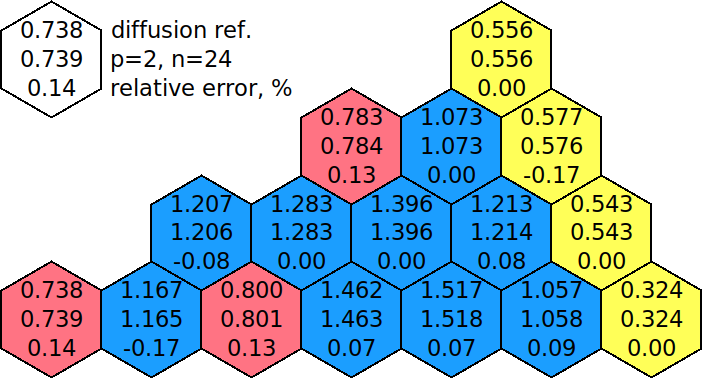}\\
	\caption{Power and error distributions using the diffusion model.}
	\label{fig:power_iaea_without_dif}
\end{center}
\end{figure}

\begin{figure}[H]
\begin{center}
	\includegraphics[width=0.75\linewidth]{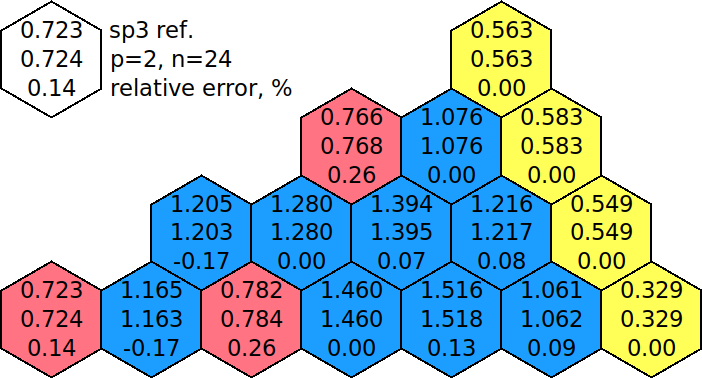}\\
	\caption{Power and error distributions using the $\mathrm{SP_3}$ model.}
	\label{fig:power_ieae_without_sp3}
\end{center}
\end{figure}

The results of the first 10 eigenvalues for $ p = 3, \  n = 96 $ are shown in Table~\ref{tab:iaea_without_lambda_10}.
The power and error distributions for the diffusion and $\mathrm{SP_3}$ models are presented in Figs \ref{fig:power_iaea_without_dif} and \ref{fig:power_ieae_without_sp3} for $p = 2, \ n = 24$.
Hereinafter, for each assembly the following data are given: the reference solution (the diffusion or $\mathrm{SP_3}$ model), the solution for $p = 2, \ n = 24$ and the relative error from the reference solution.

\begin{table}[h]
\caption{The effective multiplication factor.}
\label{tab:iaea_without_lambda}
\begin{center}
\begin{tabular}{c r r r r r r r}
\hline
$n$ & $p$ & $k_{dif}$ & $\Delta_{dif},pcm$ & $\delta_{dif}$ &$k_{sp_3}$& $\Delta_{sp_3},pcm$ & $\delta_{sp_3}$ \\
\hline
	& 1	& 0.97335& 473& 3.80& 0.97445& 490&  4.02\\
6	& 2	& 0.97760&  48& 0.45& 0.97881&  54&  0.52\\
	& 3	& 0.97801&   7& 0.07& 0.97925&  10&  0.09\\
\hline
	& 1	& 0.97654& 154& 1.28& 0.97772& 163& 1.38\\
24& 2	& 0.97799&   9& 0.08& 0.97923&  12& 0.11\\
	& 3	& 0.97807&   1& 0.01& 0.97934&   1& 0.02\\ 
\hline
	& 1	& 0.97765&  43& 0.36& 0.97888&  47& 0.40\\
96& 2	& 0.97807&   1& 0.02& 0.97933&   2& 0.02\\
	& 3	& 0.97808&   0& 0.01& 0.97935&  --& --\\ 
\hline
Ref.&   & 0.97808&    &     & 0.97935&    &\\ 
\hline
\end{tabular}
\end{center}
\end{table}

\begin{table}[h]
\caption{The eigenvalues $k_i=1/\lambda_i^{(k)}$ for $p=3, n=96$.}
\label{tab:iaea_without_lambda_10}
\begin{center}
\begin{tabular}{c r r}
\hline
$i$ & Diffusion & SP$_3$  \\
\hline
1 & 0.97808 + 0.0$i$ & 0.979351 + 0.0$i$\\
2 & 0.96318 + 0.0$i$ & 0.964604 + 0.0$i$\\
3 & 0.96318 + 0.0$i$ & 0.964604 + 0.0$i$\\
4 & 0.93844 + 0.0$i$ & 0.940253 + 0.0$i$\\
5 & 0.93844 + 0.0$i$ & 0.940253 + 0.0$i$\\
6 & 0.91966 + 0.0$i$ & 0.921844 + 0.0$i$\\
7 & 0.90220 + 0.0$i$ & 0.904467 + 0.0$i$\\
8 & 0.87141 + 0.0$i$ & 0.874997 + 0.0$i$\\
9 & 0.84957 + 0.0$i$ & 0.853155 + 0.0$i$\\
10 & 0.84957 + 0.0$i$ & 0.853154 + 0.0$i$\\
\hline
\end{tabular}
\end{center}
\end{table}


\subsubsection{Solution of $\alpha$-spectral problem without delayed neutrons}

As a reference solution for both the diffusion and transport $\mathrm{SP_3}$ models, we used the solutions obtained using very fine mesh ($p = 3, \  n = 96$).
Hereinafter, for $\alpha$-spectral problems, the following notation is used: $\alpha_{dif}$ --- $\alpha$-eigenvalue by diffusion model; $\alpha_{sp_3}$ --- $\alpha$-eigenvalue by $\mathrm{SP_3}$ model; $\Delta$ --- absolute deviation from the reference value.

The calculation results for the $\alpha$-spectral problem without  delayed neutrons using the different meshes and the finite element orders are shown in Table~\ref{tab:iaea_without_alpha}.
These data demonstrate the convergence of approximate computed eigenvalues with refinement of the calculation mesh and increase in polynomial degree.

\begin{table}[h]
\caption{The $\alpha$- eigenvalues.}
\label{tab:iaea_without_alpha}
\begin{center}
\begin{tabular}{c c r r r r}
\hline
$n$ & $p$ & $\alpha_{dif}$ & $\Delta_{dif}$ &$\alpha_{sp_3}$& $\Delta_{sp_3}$ \\
\hline
	& 1	& 556.3 & 100.8 & 532.7 & 104.1\\
6	& 2	& 465.6 & 10.1 & 440.0 & 11.4\\
	& 3	& 457.0 &  1.5 & 430.7 & 2.1\\ 
\hline
	& 1	& 488.1 & 32.6 & 463.0 & 34.4\\
24& 2	& 457.4 & 1.9 & 431.0 & 2.4\\
	& 3	& 455.7 & 0.2 & 428.9 & 0.3\\ 
\hline
	& 1	& 464.6 & 9.1 & 438.4 & 9.8\\
96& 2	& 455.8 & 0.3 & 428.9 & 0.3\\
	& 3	& 455.5 & -- & 428.6 & -- \\ 
\hline
Ref.& & 455.5 & & 428.6 \\ 
\hline
\end{tabular}
\end{center}
\end{table}

The spectral problem results for the first 10 eigenvalues are shown in Table~\ref{tab:iaea_without_alpha_10}.
The eigenvalues $\lambda_1^{(\alpha)} \leq \lambda_2^{(\alpha)} \leq ...$ are well separated. 
In this example, the fundamental eigenvalue is less compared the rest and therefore the main harmonic  will attenuate more slowly.
A regular mode of the reactor is thereby defined.
The value $\alpha = \lambda_1^{(\alpha)}$ determines the amplitude of neutron flux and is connected directly with reactor period in the regular mode.

\begin{table}[h]
\caption{The eigenvalues $\alpha_i=\lambda_i^{(\alpha)}$ for $p=3, n=96$.}
\label{tab:iaea_without_alpha_10}
\begin{center}
\begin{tabular}{c r r}
\hline
$i$ & Diffusion & SP$_3$ \\
\hline
1 & 455.540 + 0.0$i$& 428.561 + 0.0$i$ \\
2 & 760.532 + 0.0$i$& 730.398 + 0.0$i$ \\
3 & 760.543 + 0.0$i$& 730.408 + 0.0$i$ \\
4 & 1267.192 + 0.0$i$&1228.835 + 0.0$i$ \\
5 & 1267.192 + 0.0$i$&1228.836 + 0.0$i$ \\
6 & 1647.145 + 0.0$i$&1601.437 + 0.0$i$ \\
7 & 2083.289 + 0.0$i$&2031.778 + 0.0$i$ \\
8 & 2696.887 + 0.0$i$&2616.862 + 0.0$i$ \\
9 & 3188.356 + 0.0$i$&3092.715 + 0.0$i$ \\
10& 3188.363 + 0.0$i$&3092.722 + 0.0$i$ \\
\hline
\end{tabular}
\end{center}
\end{table}

\begin{figure}[H]
\begin{center}
	\includegraphics[width=0.49\linewidth]{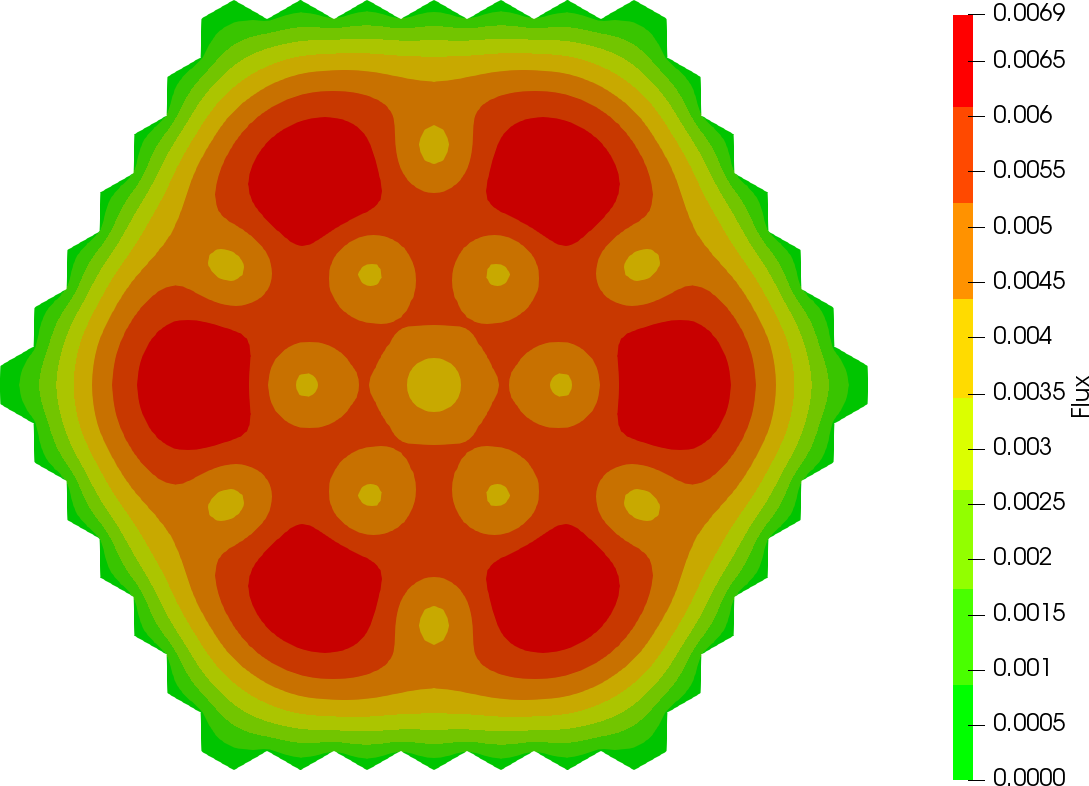}
	\includegraphics[width=0.49\linewidth]{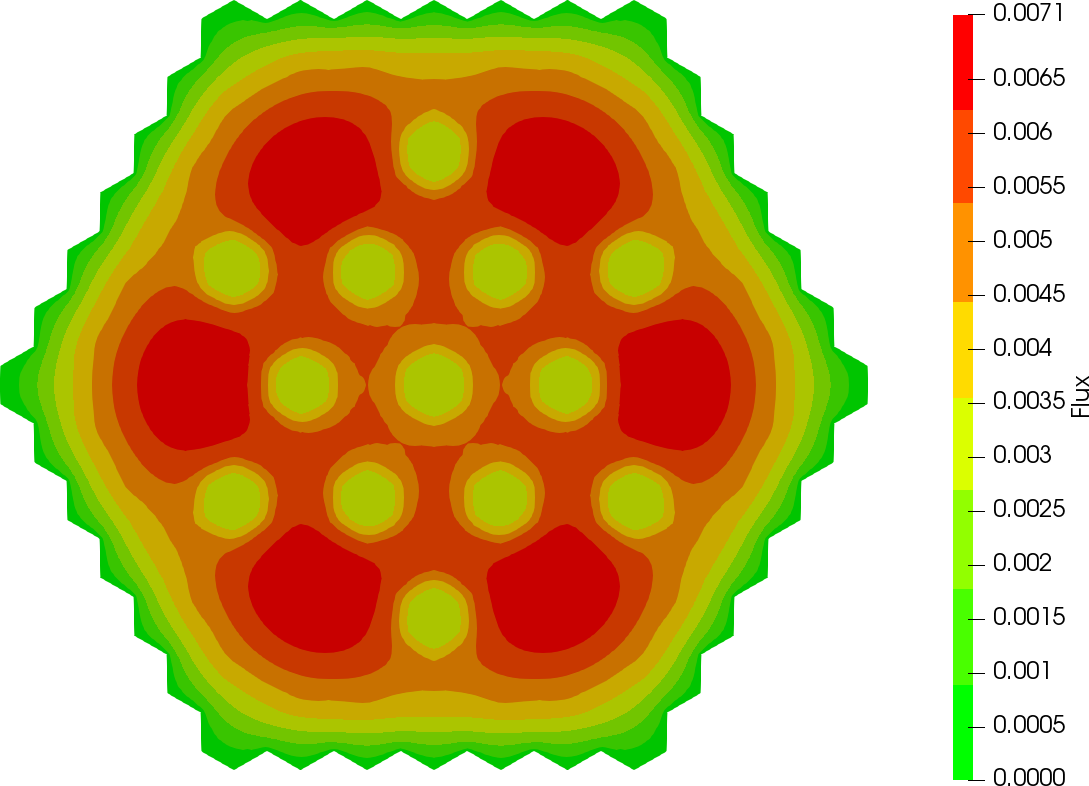}\\
	\caption{Eigenfunctions $\phi_1^{(1)}$, $\phi_2^{(1)}$ using $\mathrm{SP_3}$ model.}
	\label{fig:iaea_without_fun_1}
\end{center}
\end{figure}

\begin{figure}[H]
\begin{center}
	\includegraphics[width=0.49\linewidth]{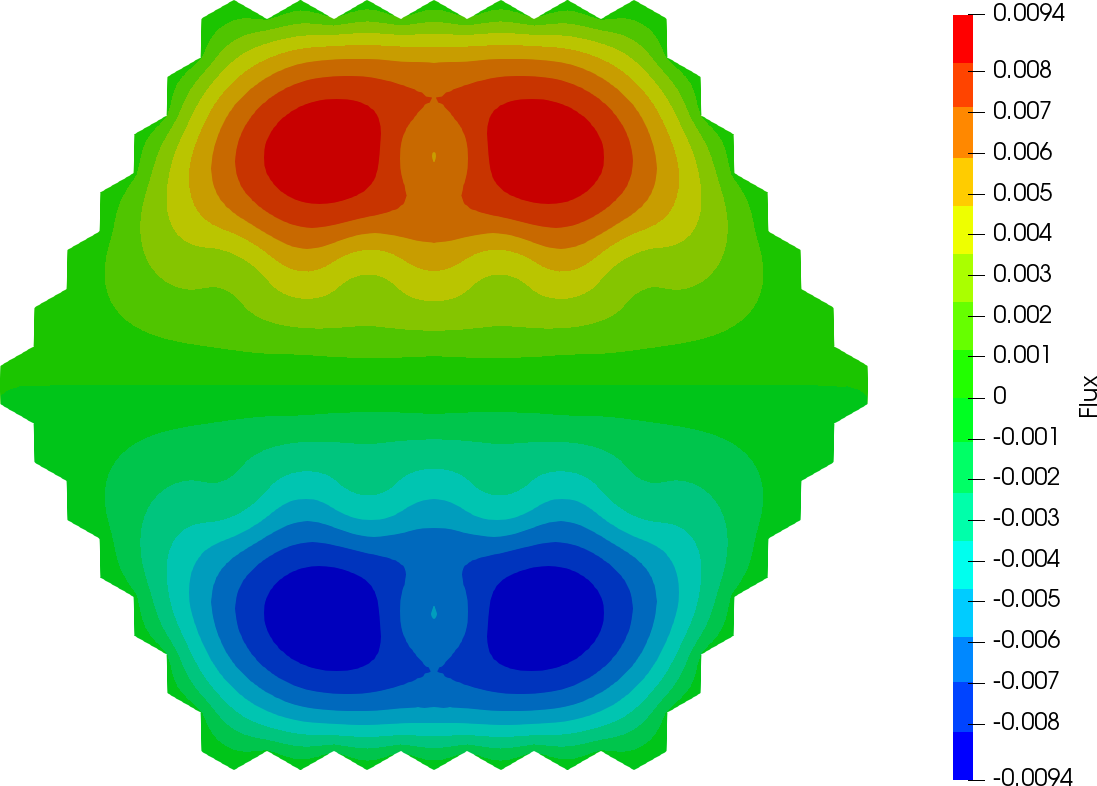}
	\includegraphics[width=0.49\linewidth]{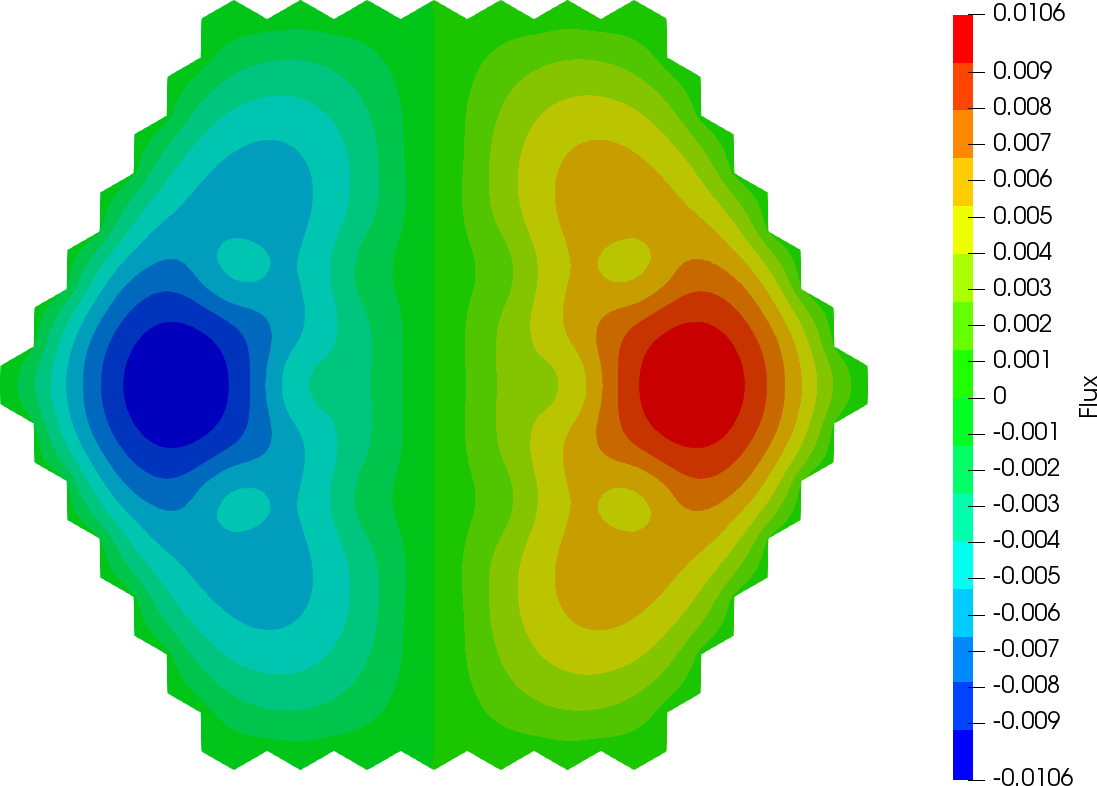}\\
	\caption{Eigenfunctions $\phi_1^{(2)}$, $\phi_1^{(3)}$ using $\mathrm{SP_3}$ model.}
	\label{fig:iaea_without_fun_2}
\end{center}
\end{figure}

\begin{figure}[H]
\begin{center}
	\includegraphics[width=0.49\linewidth]{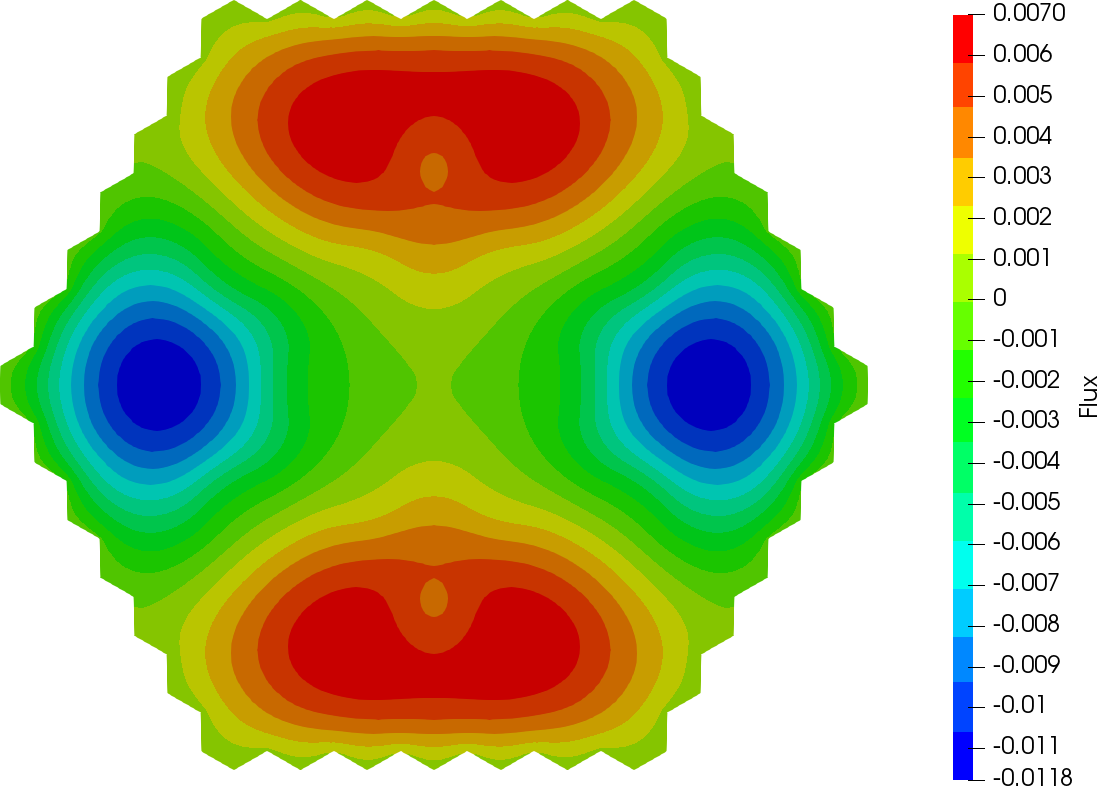}
	\includegraphics[width=0.49\linewidth]{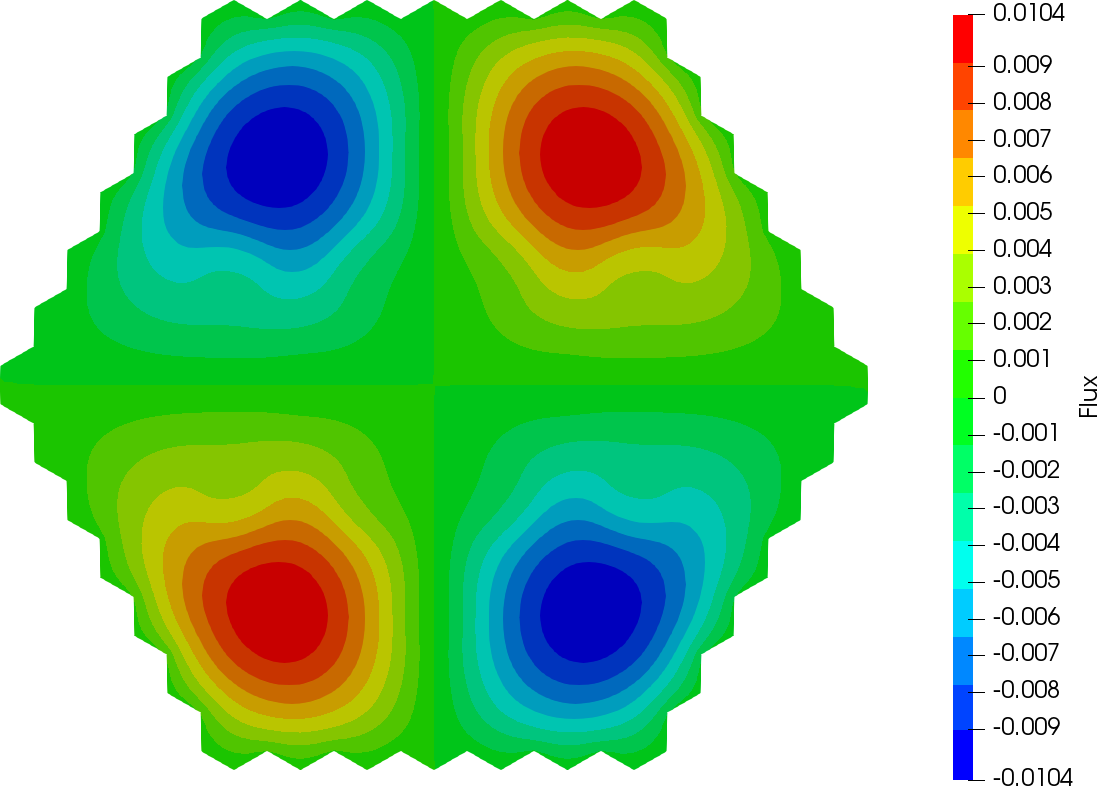}\\
	\caption{Eigenfunctions $\phi_1^{(4)}$, $\phi_1^{(5)}$ using $\mathrm{SP_3}$ model.}
	\label{fig:iaea_without_fun_3}
\end{center}
\end{figure}

The eigenfunctions for fundamental eigenvalue ($i=1$) of the $\alpha$-spectral problem without delayed neutron are shown in Fig.~\ref{fig:iaea_without_fun_1}. 
Due to the fact that the reactor state is close to critical ($k=k_1\approx 0.97935$), the fundamental eigenfunctions of the $\lambda$-spectral problem are close to the fundamental eigenfunctions of the $\alpha$-spectral problem.
The eigenfunctions $\phi_1^{(i)}, i=2,3,4,5$ are shown in Fig.~\ref{fig:iaea_without_fun_2}, Fig.~\ref{fig:iaea_without_fun_3}.


\subsubsection{Solution of $\alpha$-spectral problem with delayed neutrons}

As a reference solution for both the diffusion and transport $\mathrm{SP_3}$ models, we used the solutions obtained using very fine mesh ($p = 3, \  n = 96$). 
The  $\alpha$-spectral problem results with delayed neutrons using the different meshes and finite element orders are shown in Table~\ref{tab:iaea_without_alpha_del}.
Compared with the previous case without delayed neutrons, these data demonstrate the similar convergence of the computed eigenvalues.

\begin{table}[h]
\caption{The $\alpha$-eigenvalues.}
\label{tab:iaea_without_alpha_del}
\begin{center}
\begin{tabular}{c c r r r r}
\hline
$n$ & $p$ & $\alpha_{dif}$ & $\Delta_{dif}$ &$\alpha_{sp_3}$& $\Delta_{sp_3}$ \\
\hline
	& 1	&0.06465 &0.00264&0.06410 & 0.00295\\
6	& 2	&0.06232 &0.00031&0.06153 & 0.00035\\
	& 3	&0.06206 &0.00005&0.06122 & 0.00007\\ 
\hline
	& 1	&0.06296 &0.00095&0.06224 & 0.00109\\
24& 2	&0.06207 &0.00005&0.06123 & 0.00008\\
	& 3	&0.06202 &0.00001&0.06116 & 0.00001\\ 
\hline
	& 1	&0.06228 &0.00027&0.06147 & 0.00032\\
96& 2	&0.06202 &0.00001&0.06116 & 0.00001\\
	& 3	&0.06201 &     --&0.06115 &      --\\ 
\hline
Ref.& & 0.06201 & & 0.06115 \\ 
\hline
\end{tabular}
\end{center}
\end{table}

\begin{table}[h]
\caption{The eigenvalues $\alpha_i=\lambda_i^{(\alpha)}$ for $p=3, n=96$.}
\label{tab:iaea_without_alpha_del_10}
\begin{center}
\begin{tabular}{c r r}
\hline
$i$ & Diffusion & SP$_3$ \\
\hline
1& 0.06201 + 0.0$i$&0.06115 + 0.0$i$\\
2& 0.06837 + 0.0$i$&0.06796 + 0.0$i$\\
3& 0.06837 + 0.0$i$&0.06796 + 0.0$i$\\
4& 0.07279 + 0.0$i$&0.07258 + 0.0$i$\\
5& 0.07279 + 0.0$i$&0.07258 + 0.0$i$\\
6& 0.07446 + 0.0$i$&0.07430 + 0.0$i$\\
7& 0.07547 + 0.0$i$&0.07536 + 0.0$i$\\
8& 0.07662 + 0.0$i$&0.07652 + 0.0$i$\\
9& 0.07717 + 0.0$i$&0.07709 + 0.0$i$\\
10& 0.07721 + 0.0$i$&0.07711 + 0.0$i$\\
\hline
\end{tabular}
\end{center}
\end{table}

The first 10 spectral problem eigenvalues are shown in Table~\ref{tab:iaea_without_alpha_del_10}.
Due to the contribution of delayed neutrons, the fundamental eigenvalue is much smaller compared with the case without delayed neutrons.

\begin{figure}[H]
\begin{center}
	\includegraphics[width=0.49\linewidth]{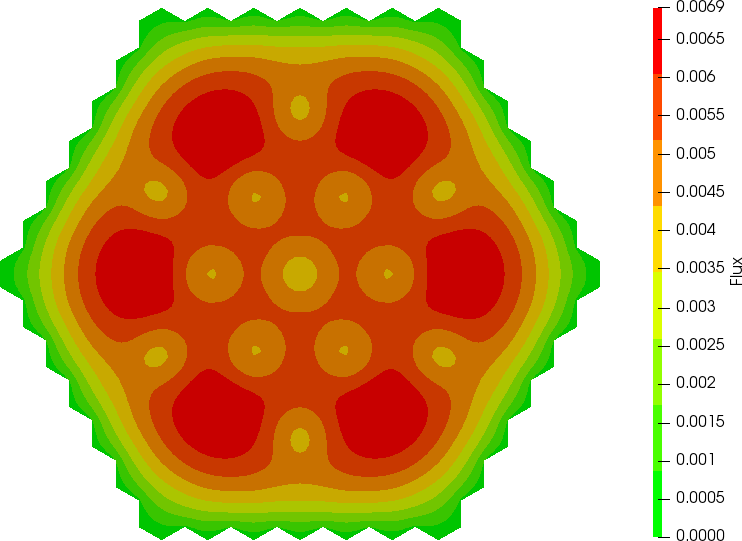}
	\includegraphics[width=0.49\linewidth]{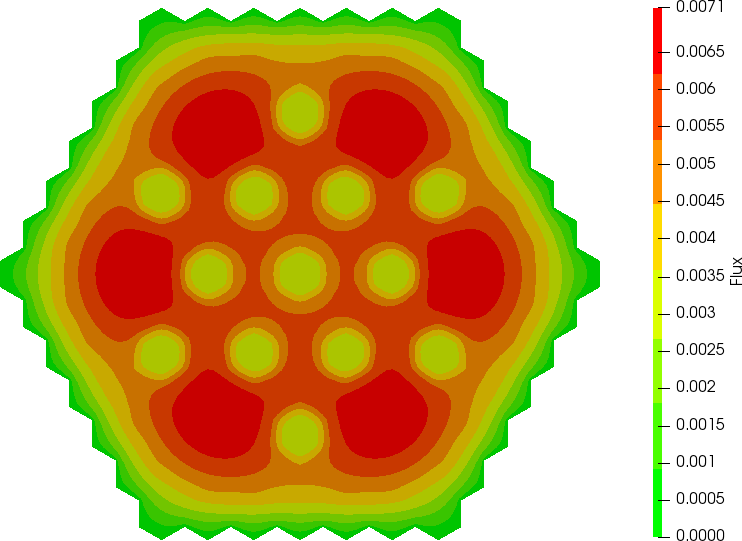}\\
	\caption{Eigenfunctions $\phi_1^{(1)}$, $\phi_2^{(1)}$.}
	\label{fig:iaea_without_fun_del_1}
\end{center}
\end{figure}

\begin{figure}[H]
\begin{center}
	\includegraphics[width=0.49\linewidth]{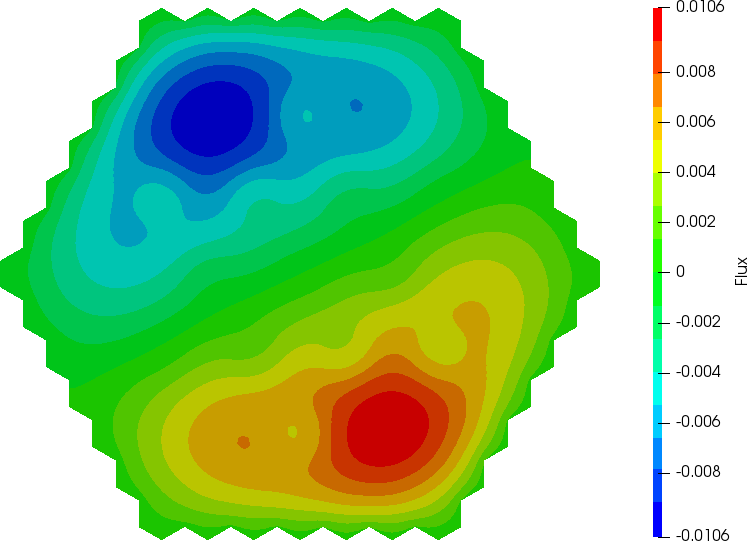}
	\includegraphics[width=0.49\linewidth]{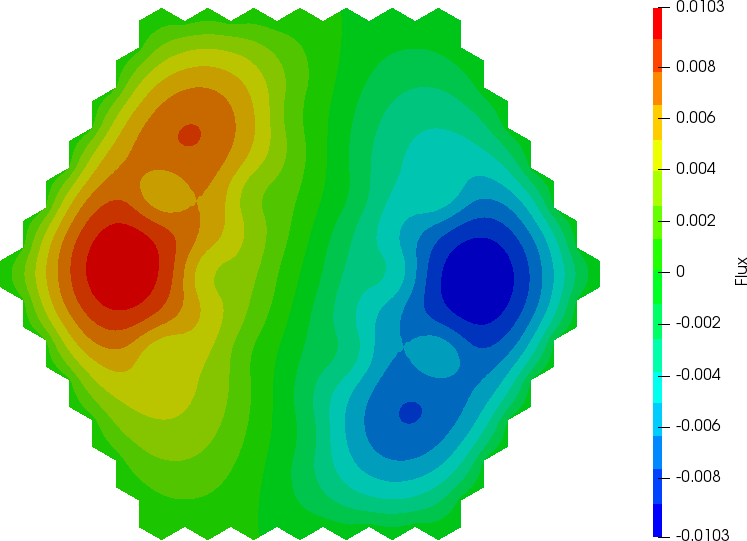}\\
	\caption{Eigenfunctions $\phi_1^{(2)}$, $\phi_1^{(3)}$.}
	\label{fig:iaea_without_fun_del_2}
\end{center}
\end{figure}

\begin{figure}[H]
\begin{center}
	\includegraphics[width=0.49\linewidth]{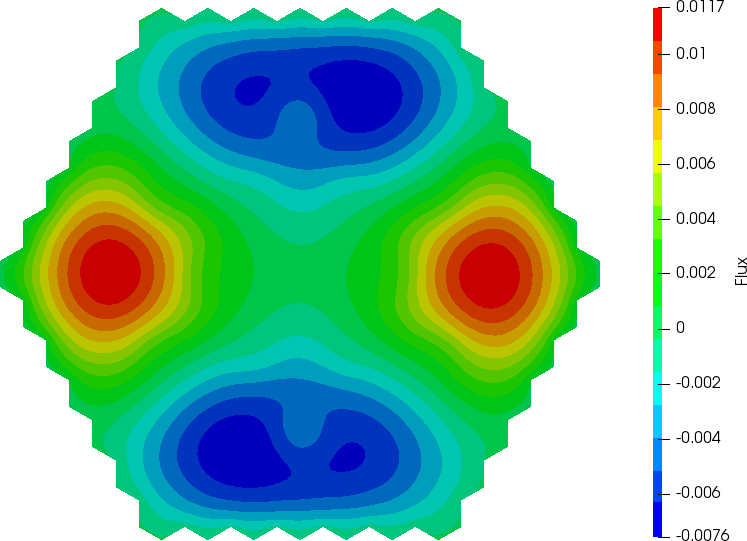}
	\includegraphics[width=0.49\linewidth]{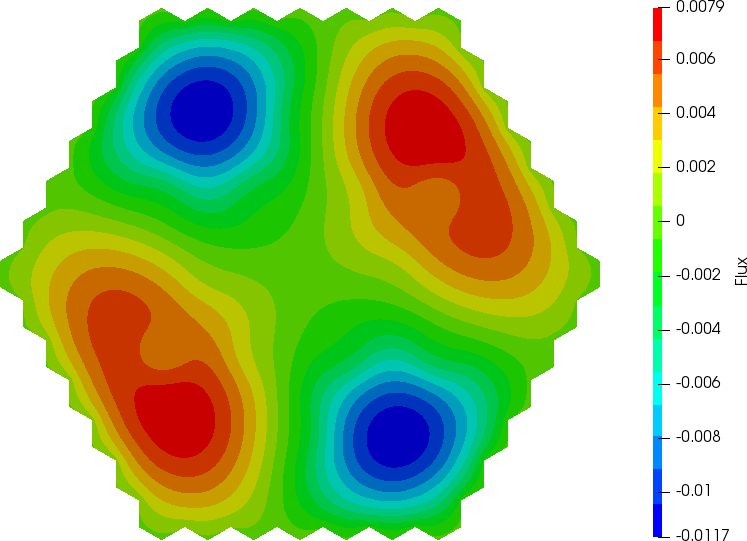}\\
	\caption{Eigenfunctions $\phi_1^{(4)}$, $\phi_1^{(5)}$.}
	\label{fig:iaea_without_fun_del_3}
\end{center}
\end{figure}

The eigenfunctions for fundamental eigenvalue ($i=1$) of the $\alpha$-spectral problem with delayed neutron are shown in Fig.~\ref{fig:iaea_without_fun_del_1}. 
The eigenfunctions $\phi_1^{(i)}, i=2,3,4,5$ are shown in Fig.~\ref{fig:iaea_without_fun_del_2}, Fig.~\ref{fig:iaea_without_fun_del_3}.
The eigenfunctions of the problems without and with delayed neutrons are close to each other in topology.


According to (\ref{1.11})-(\ref{1.13}), the prompt neutron generation time $\Lambda_{pr} = 4.920 \cdot 10^{-5} s$  for both the diffusion and $\mathrm{SP_3}$ fundamental eigenvalues.
Assuming the same value of $\Lambda_{pr}$ for both $ \alpha $-spectral problems, the uncertainty in the fundamental eigenvalues is less than 0.1 percent.

\subsection{IAEA-2D with reflector}

This test differs from the previous one only additional external row of reflector assemblies (material 4, see Table~\ref{tab:iaea}). 

\subsubsection{Solution of Lambda Modes spectral problem}

As a reference solution for the diffusion model, we used the previous results \citep{avvakumov2015}; for the $\mathrm{SP_3}$ model --- the solution obtained using the MCNP4C code \citep{Bahabadi2016}. As well as in the previous benchmark calculations, the maximum difference in assembly power between two models is about 2 percent for the rodded assemblies (material 3, see Fig.~\ref{fig:iaea}).

The comparison of the calculated effective multiplication factors is shown in Table~\ref{tab:iaea_with_lambda}. 
The results of the first 10 eigenvalues for $ p = 3, \  n = 96 $ are presented in Table~\ref{tab:iaea_with_lambda_10}.
One can see that there are several eigenvalues of multiplicity two.

\begin{figure}[H]
\begin{center}
	\includegraphics[width=0.75\linewidth]{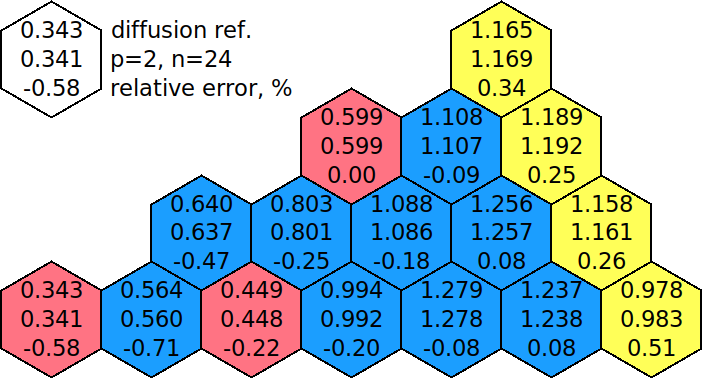}\\
	\caption{Power and error distributions using the diffusion model.}
	\label{fig:power_iaea_with_dif}
\end{center}
\end{figure}

\begin{figure}[H]
\begin{center}
	\includegraphics[width=0.75\linewidth]{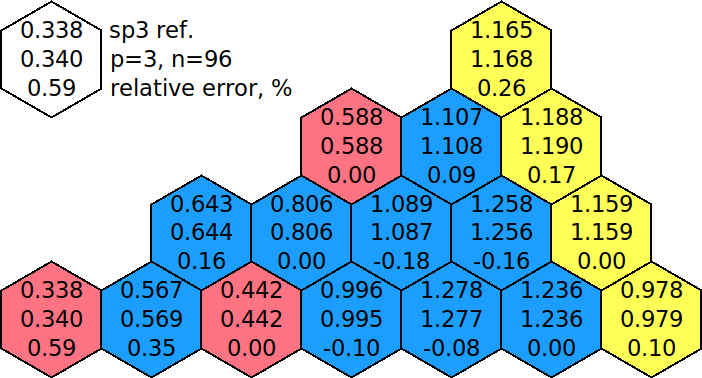}\\
	\caption{Power and error distributions using the $\mathrm{SP_3}$ model.}
	\label{fig:power_iaea_with_sp3}
\end{center}
\end{figure}

The power distributions and calculation errors for $p = 2, n = 24$ using the diffusion model are shown in Fig~\ref{fig:power_iaea_with_dif} and for $p = 3, \  n = 96$ using the $\mathrm{SP_3}$ model are shown Fig~\ref{fig:power_iaea_with_sp3}.

\begin{table}[h]
\caption{The effective multiplication factor.}
\label{tab:iaea_with_lambda}
\begin{center}
\begin{tabular}{c c r r r r r r}
\hline
$n$ & $p$ & $k_{dif}$ & $\Delta_{dif},pcm$ & $\delta_{dif}$ &$k_{sp_3}$& $\Delta_{sp_3},pcm$ & $\delta_{sp_3}$ \\
\hline
	& 1	& 1.01041& 490&13.29& 1.01159& 536& 14.14\\
6	& 2	& 1.00623&  72& 1.88& 1.00711&  88&  2.19\\
	& 3	& 1.00558&   7& 0.22& 1.00636&  13&  0.35\\ 
\hline
	& 1	& 1.00699& 148& 4.54& 1.00792& 169&  4.96\\
24& 2	& 1.00561&  10& 0.30& 1.00640&  17&  0.42\\
	& 3	& 1.00551&   0& 0.02& 1.00626&   3&  0.17\\ 
\hline
	& 1	& 1.00591&  36& 1.28& 1.00671&  48&  1.42\\
96& 2	& 1.00552&   1& 0.04& 1.00626&   3&  0.18\\
	& 3	& 1.00551&   0& 0.01& 1.00625&   2&  0.18\\ 
\hline
Ref.&   & 1.00551&    &     & 1.00623&     &\\ 
\hline
\end{tabular}
\end{center}
\end{table}

\begin{table}[h]
\caption{The eigenvalues $k_i=1/\lambda_i^{(k)}$ for $p=3, n=96$.}
\label{tab:iaea_with_lambda_10}
\begin{center}
\begin{tabular}{c r r}
\hline
$i$ & Diffusion & SP$_3$  \\
\hline
1 & 1.005510 + 0.0$i$   & 1.006245 + 0.0$i$\\
2 & 0.996490 + 0.0$i$ & 0.997254 + 0.0$i$\\
3 & 0.996490 + 0.0$i$ & 0.997254 + 0.0$i$\\
4 & 0.976791 + 0.0$i$ & 0.977759 + 0.0$i$\\
5 & 0.976791 + 0.0$i$ & 0.977759 + 0.0$i$\\
6 & 0.958684 + 0.0$i$ & 0.959895 + 0.0$i$\\
7 & 0.928980 + 0.0$i$ & 0.930969 + 0.0$i$\\
8 & 0.924186 + 0.0$i$ & 0.925931 + 0.0$i$\\
9 & 0.904788 + 0.0$i$ & 0.907349 + 0.0$i$\\
10 & 0.904788 + 0.0$i$ & 0.907349 + 0.0$i$\\
\hline
\end{tabular}
\end{center}
\end{table}


\subsubsection{Solution of $\alpha$-spectral problem without delayed neutrons}

As a reference solution, we use the fine mesh solutions obtained using the diffusion or transport $\mathrm{SP_3}$ model ($ p = 3, \  n = 96 $). 
The $\alpha$-spectral problem results are shown in Table~\ref{tab:iaea_with_alpha}.

\begin{table}[h]
\caption{The $\alpha$-eigenvalues.}
\label{tab:iaea_with_alpha}
\begin{center}
\begin{tabular}{c c r r r r}
\hline
$n$ & $p$ & $\alpha_{dif}$ & $\Delta_{dif}$ &$\alpha_{sp_3}$& $\Delta_{sp_3}$ \\
\hline
	& 1	& $-$184.95 & 84.14 & $-$205.92 & 91.32\\
6	& 2	& $-$113.58 & 12.77 & $-$130.02 & 15.42\\
	& 3	& $-$101.98 &  1.17 & $-$116.72 &  2.12\\ 
\hline
	& 1	& $-$126.66 & 25.85 & $-$143.85 & 29.25\\
24& 2	& $-$102.58 &  1.77 & $-$117.31 &  2.71\\
	& 3	& $-$100.88 &  0.07 & $-$114.83 &  0.23\\ 
\hline
	& 1	& $-$107.82 &  7.01 & $-$122.84 & 8.24\\
96& 2	& $-$100.97 &  0.16 & $-$114.94 & 0.34\\
	& 3	& $-$100.81 &	 -- & $-$114.60 &  -- \\ 
\hline
Ref.& & $-$100.81 & & $-$114.60 \\ 
\hline
\end{tabular}
\end{center}
\end{table}

The results of the first 10 eigenvalues for $ p = 3, \  n = 96 $ are presented in Table~\ref{tab:iaea_with_lambda_10}.
As before, the eigenvalues are well separated.
In this example, the fundamental eigenvalue is negative and therefore the main harmonic will increase, while all others will attenuate. 

\begin{table}[h]
\caption{The eigenvalues $\alpha_i=\lambda_i^{(\alpha)}$ for $p=3, n=96$.}
\label{tab:iaea_with_alpha_10}
\begin{center}
\begin{tabular}{c r r}
\hline
$i$ & Diffusion & SP$_3$ \\
\hline
1 &$-$100.81 + 0.0$i$&$-$114.60 + 0.0$i$ \\
2 &  62.93 + 0.0$i$& 49.42 + 0.0$i$ \\
3 &  62.93 + 0.0$i$& 49.42 + 0.0$i$ \\
4 & 405.31 + 0.0$i$&390.15 + 0.0$i$ \\
5 & 405.31 + 0.0$i$&390.15 + 0.0$i$ \\
6 & 710.64 + 0.0$i$&693.47 + 0.0$i$ \\
7 &1141.43 + 0.0$i$&1118.67 + 0.0$i$ \\
8 &1469.68 + 0.0$i$&1438.31 + 0.0$i$ \\
9 &1494.37 + 0.0$i$&1468.54 + 0.0$i$ \\
10&1494.37 + 0.0$i$&1468.54 + 0.0$i$ \\
\hline
\end{tabular}
\end{center}
\end{table}

\begin{figure}[H]
\begin{center}
	\includegraphics[width=0.49\linewidth]{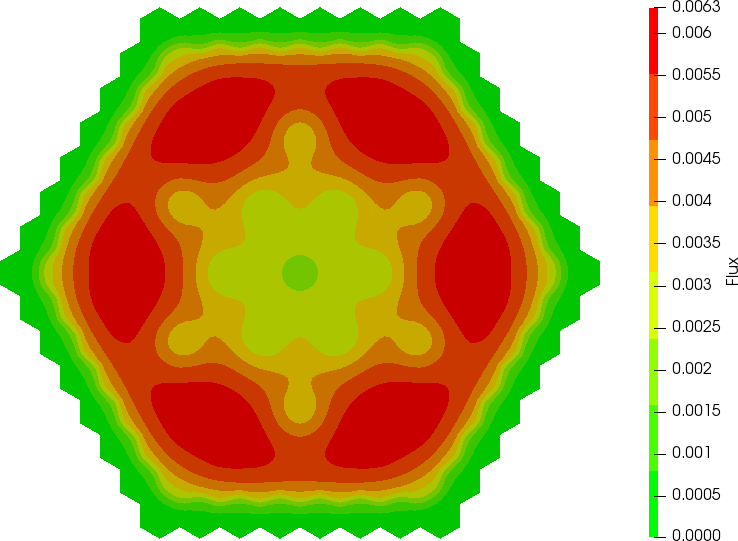}
	\includegraphics[width=0.49\linewidth]{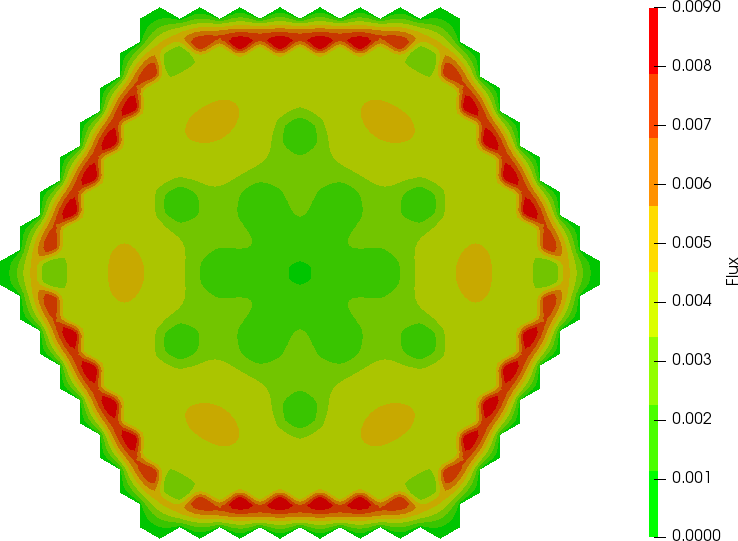}\\
	\caption{Eigenfuncions $\phi_1^{(1)}$, $\phi_2^{(1)}$.}
	\label{fig:iaea_with_fun_1}
\end{center}
\end{figure}

\begin{figure}[H]
\begin{center}
	\includegraphics[width=0.49\linewidth]{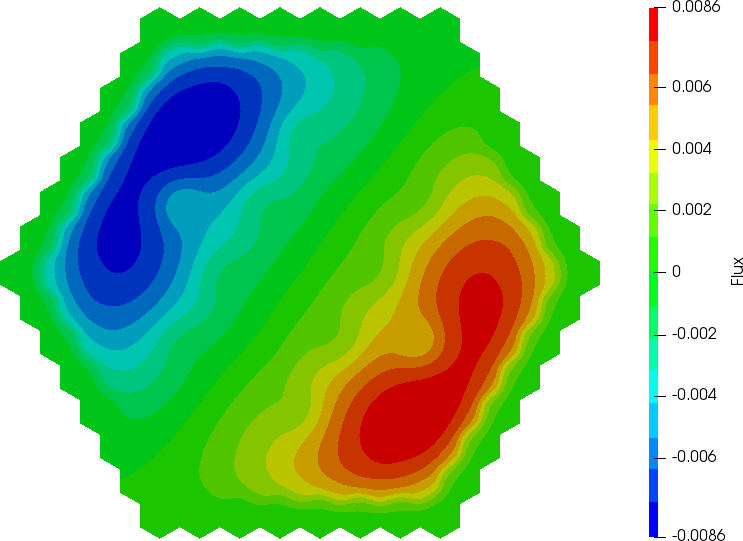}
	\includegraphics[width=0.49\linewidth]{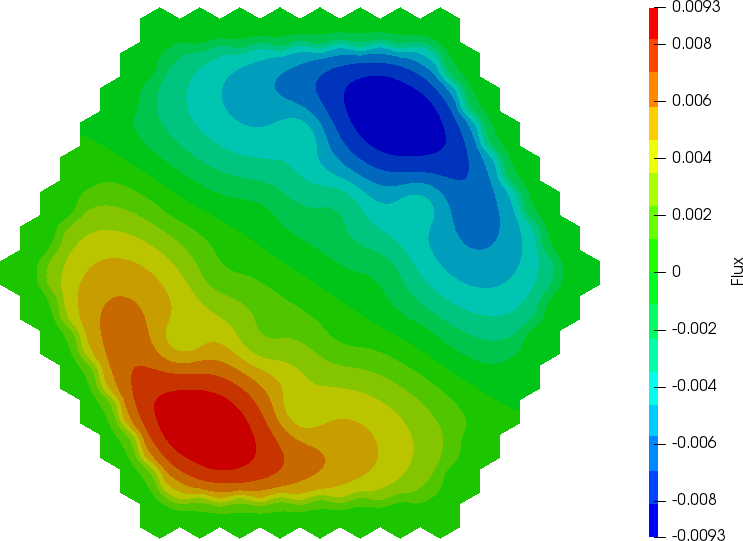}\\
	\caption{Eigenfuncions $\phi_1^{(2)}$, $\phi_1^{(3)}$.}
	\label{fig:iaea_with_fun_2}
\end{center}
\end{figure}

\begin{figure}[H]
\begin{center}
	\includegraphics[width=0.49\linewidth]{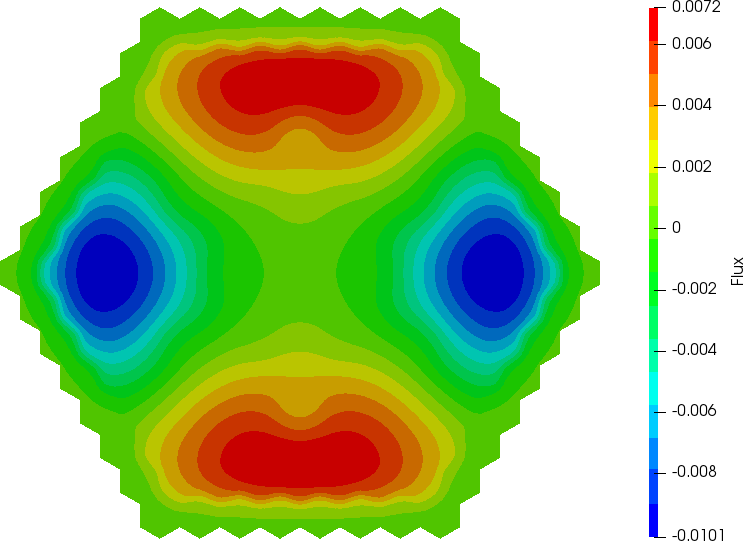}
	\includegraphics[width=0.49\linewidth]{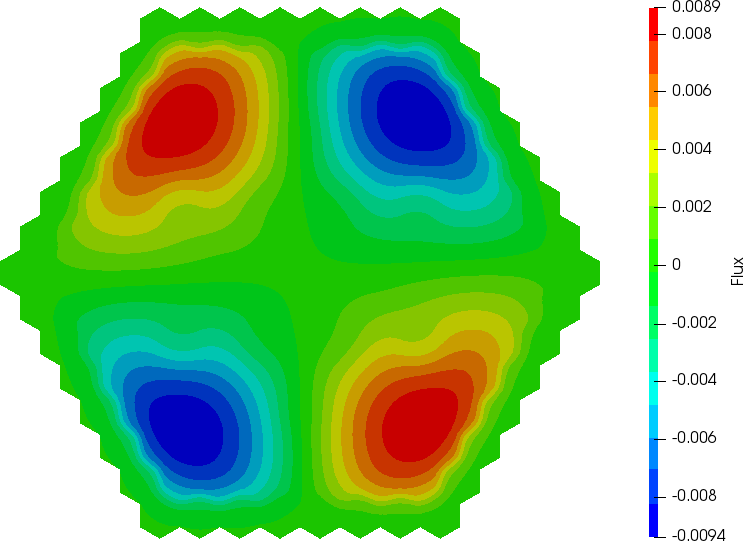}\\
	\caption{Eigenfuncions $\phi_1^{(4)}$, $\phi_1^{(5)}$.}
	\label{fig:iaea_with_fun_3}
\end{center}
\end{figure}

The eigenfunctions for fundamental eigenvalue ($i=1$) of the $\alpha$-spectral problem without delayed neutrons are shown in Fig.~\ref{fig:iaea_with_fun_1}. 
Due to the fact that a state of the reactor is close to critical ($k=k_1\approx 1.00625$), the fundamental eigenfunctions of the $\lambda$-spectral problem are close to the fundamental eigenfunctions of the $\alpha$-spectral problem.
The eigenfunctions $\phi_1^{(i)}, i=2,3,4,5$ are shown in Fig.~\ref{fig:iaea_with_fun_2}, Fig.~\ref{fig:iaea_with_fun_3}.


\subsubsection{Solution of $\alpha$-spectral problem with delayed neutrons}

As a reference solution, we use the fine mesh solutions obtained using the diffusion or transport $\mathrm{SP_3}$ model ($ p = 3, \  n = 96 $). 
The $\alpha$-spectral problem results are shown in
Table~\ref{tab:iaea_with_alpha_del}.

\begin{table}[h]
\caption{The $\alpha$-eigenvalues.}
\label{tab:iaea_with_alpha_del}
\begin{center}
\begin{tabular}{c c r r r r}
\hline
$n$ & $p$ & $\alpha_{dif}$ & $\Delta_{dif}$ &$\alpha_{sp_3}$& $\Delta_{sp_3}$ \\
\hline
	& 1	&$-$68.2268 &67.8084& $-$88.9461 &87.6086\\
6	& 2	& $-$1.2810 & 0.8626& $-$11.1554 & 9.8179\\
	& 3	& $-$0.4506 & 0.0322&  $-$1.8063 & 0.4688\\ 
\hline
	& 1	& $-$9.0267  & 8.6083&$-$25.1658 &23.8283\\
24& 2	& $-$0.4686  & 0.0502& $-$1.9832 & 0.6457\\
	& 3	& $-$0.4202  & 0.0018& $-$1.3787 & 0.0412\\ 
\hline
	& 1	& $-$0.7018  & 0.2834& $-$4.9794 & 3.6419\\
96& 2	& $-$0.4225  & 0.0041& $-$1.3994 & 0.0619\\
	& 3	& $-$0.4184  &    -- & $-$1.3375 &    --\\ 
\hline
Ref.& & $-$0.4184 & & $-$1.3375 \\ 
\hline
\end{tabular}
\end{center}
\end{table}

\begin{table}[h]
\caption{The eigenvalues $\alpha_i=\lambda_i^{(\alpha)}$ for $p=3, n=96$.}
\label{tab:iaea_with_alpha_del_10}
\begin{center}
\begin{tabular}{c r r}
\hline
$i$ & Diffusion & SP$_3$ \\
\hline
1& $-$0.4184 + 0.0$i$&$-$1.3375 + 0.0$i$\\
2& 0.0281 + 0.0$i$&0.0238 + 0.0$i$\\
3& 0.0281 + 0.0$i$&0.0238 + 0.0$i$\\
4& 0.0628 + 0.0$i$&0.0622 + 0.0$i$\\
5& 0.0628 + 0.0$i$&0.0622 + 0.0$i$\\
6& 0.0695 + 0.0$i$&0.0692 + 0.0$i$\\
7& 0.0737 + 0.0$i$&0.0736 + 0.0$i$\\
8& 0.0741 + 0.0$i$&0.0740 + 0.0$i$\\
9& 0.0754 + 0.0$i$&0.0752 + 0.0$i$\\
10& 0.0763 + 0.0$i$&0.0762 + 0.0$i$\\
\hline
\end{tabular}
\end{center}
\end{table}

The calculation results for the first 10 eigenvalues are shown in Table~\ref{tab:iaea_with_alpha_del_10}.
Due to the contribution of delayed neutrons, the fundamental eigenvalue is much smaller than in the case without delayed neutrons. Again the fundamental eigenvalue is negative and therefore the main harmonic will increase, while all others will attenuate.

\begin{figure}[H]
\begin{center}
	\includegraphics[width=0.49\linewidth]{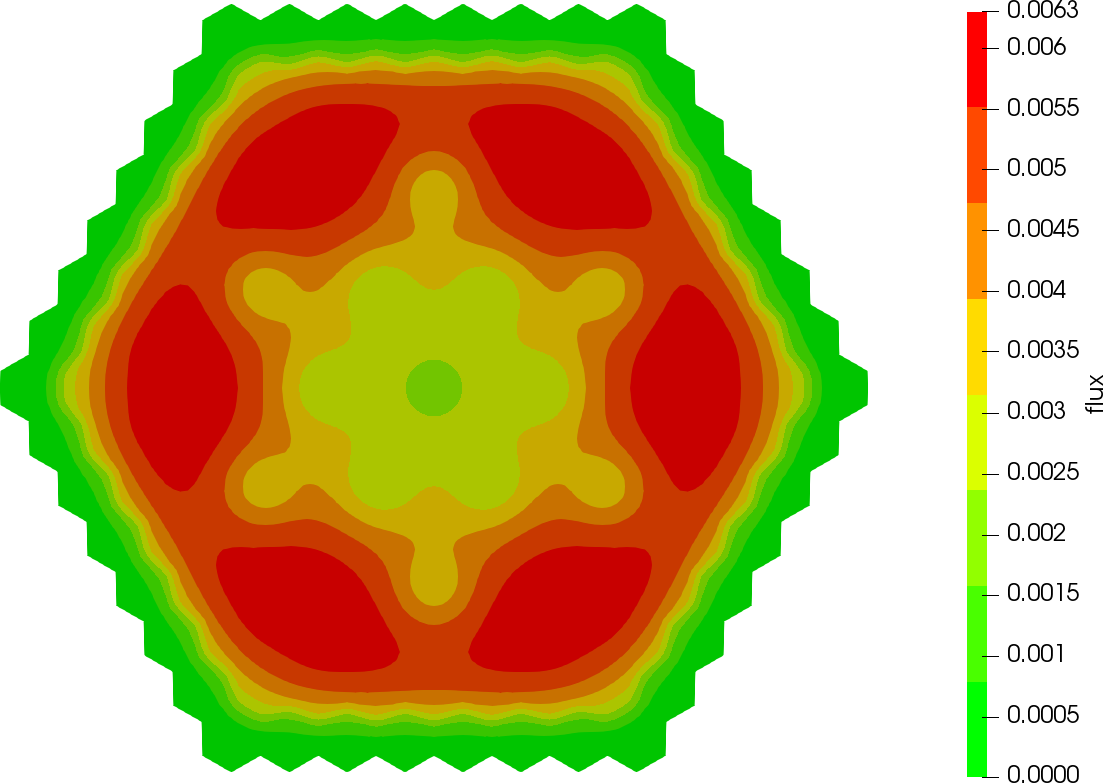}
	\includegraphics[width=0.49\linewidth]{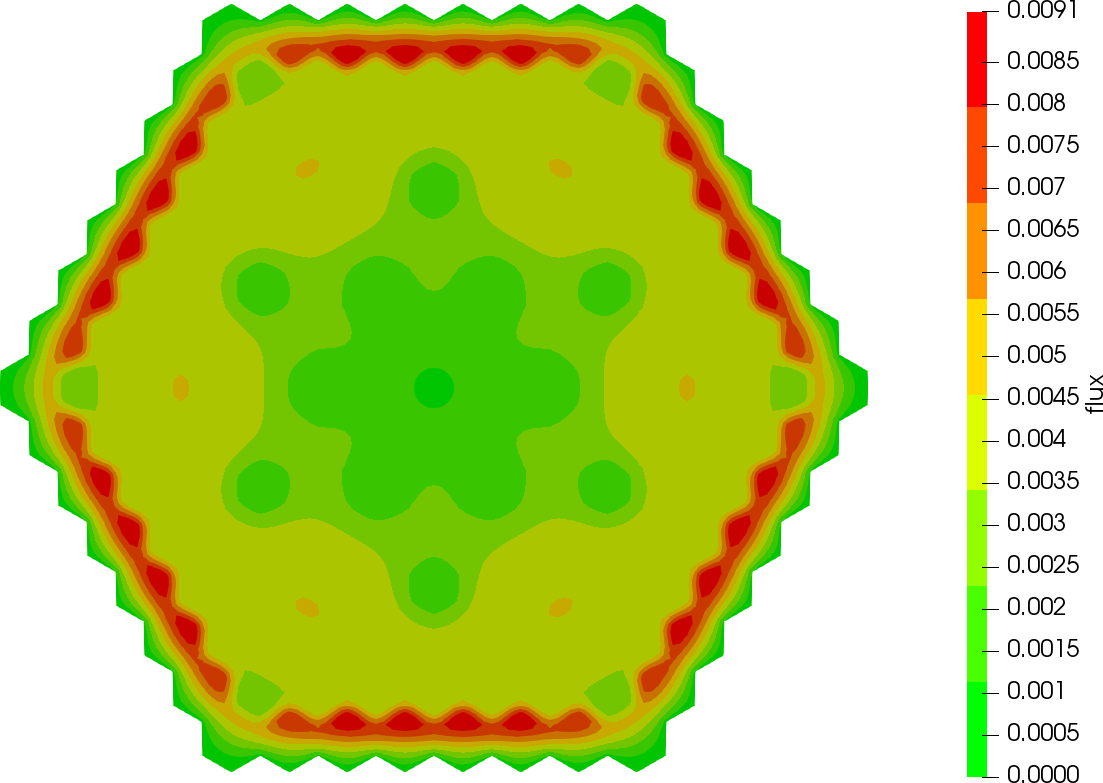}\\
	\caption{Eigenfunctions $\phi_1^{(1)}$, $\phi_2^{(1)}$.}
	\label{fig:iaea_with_fun_del_1}
\end{center}
\end{figure}

\begin{figure}[H]
\begin{center}
	\includegraphics[width=0.49\linewidth]{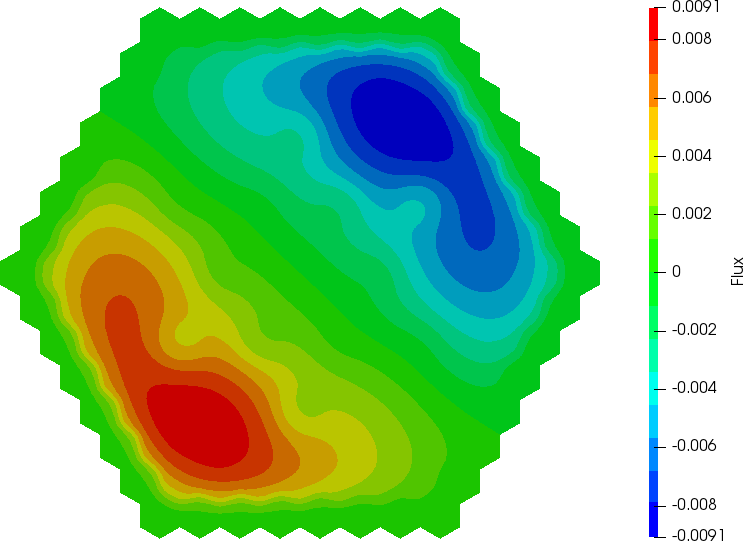}
	\includegraphics[width=0.49\linewidth]{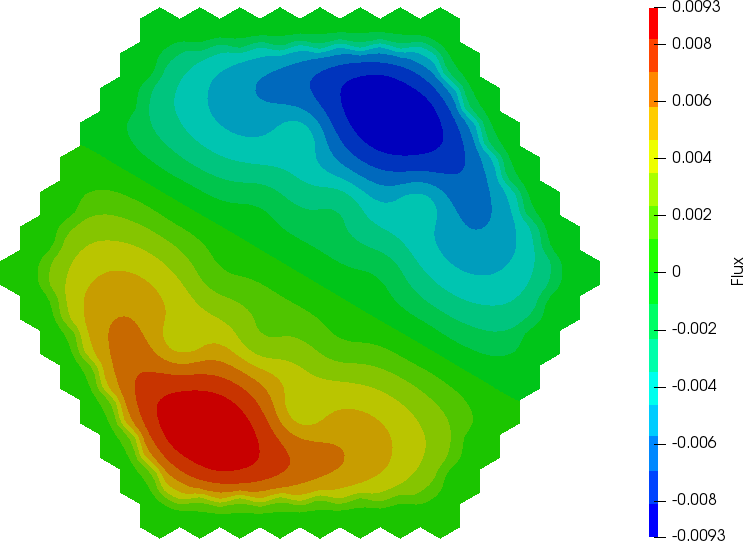}\\
	\caption{Eigenfunctions $\phi_1^{(2)}$, $\phi_1^{(3)}$.}
	\label{fig:iaea_with_fun_del_2}
\end{center}
\end{figure}

\begin{figure}[H]
\begin{center}
	\includegraphics[width=0.49\linewidth]{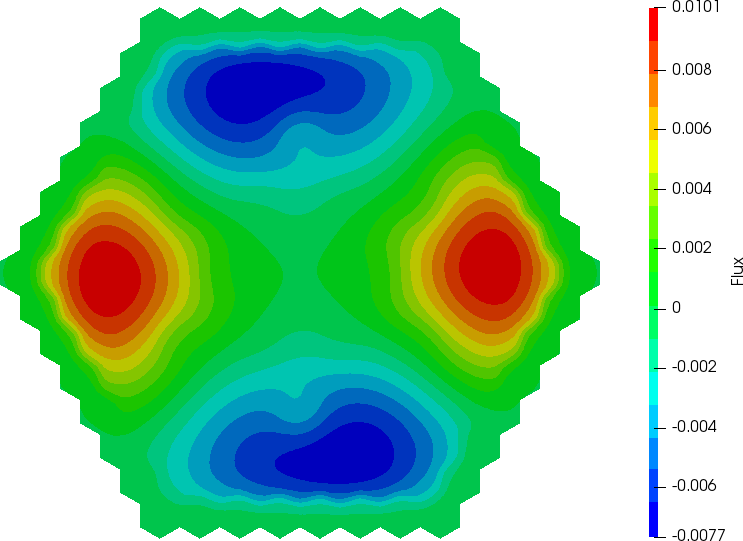}
	\includegraphics[width=0.49\linewidth]{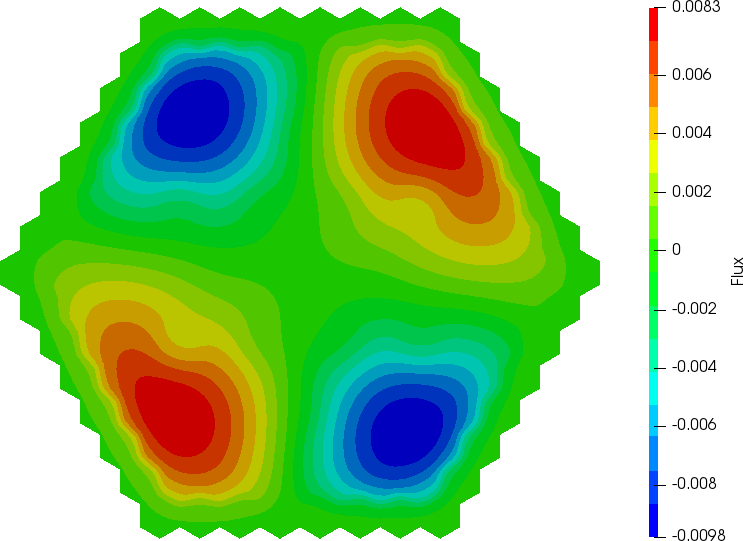}\\
	\caption{Eigenfunctions $\phi_1^{(4)}$, $\phi_1^{(5)}$.}
	\label{fig:iaea_with_fun_del_3}
\end{center}
\end{figure}

The eigenfunctions for fundamental eigenvalue ($i=1$) of the $\alpha$-spectral problem with delayed neutrons are presented in Fig.~\ref{fig:iaea_with_fun_del_1}. 
The eigenfunctions $\phi_1^{(i)}, i=2,3,4,5$ are shown in Fig.~\ref{fig:iaea_with_fun_del_2}, Fig.~\ref{fig:iaea_with_fun_del_3}.
The eigenfunctions of the problems without and with delayed neutrons are close to each other in topology.


According to (\ref{1.11})-(\ref{1.13}), the prompt neutron generation time $\Lambda_{pr} = 5.436 \cdot 10^{-5} s$  for  the diffusion fundamental eigenvalue and $\Lambda_{pr} = 5.416 \cdot 10^{-5} s$  for the $\mathrm{SP_3}$ fundamental eigenvalue. 
Assuming the same value of $\Lambda_{pr}$ for both $ \alpha $-spectral problems, the uncertainty in the fundamental eigenvalues is less than 0.1 percent.

\subsection{Azimutally non-symmetric test IAEA-2D with reflector}

\begin{figure}[h]
	\begin{center}
		\includegraphics[width=0.75\linewidth]{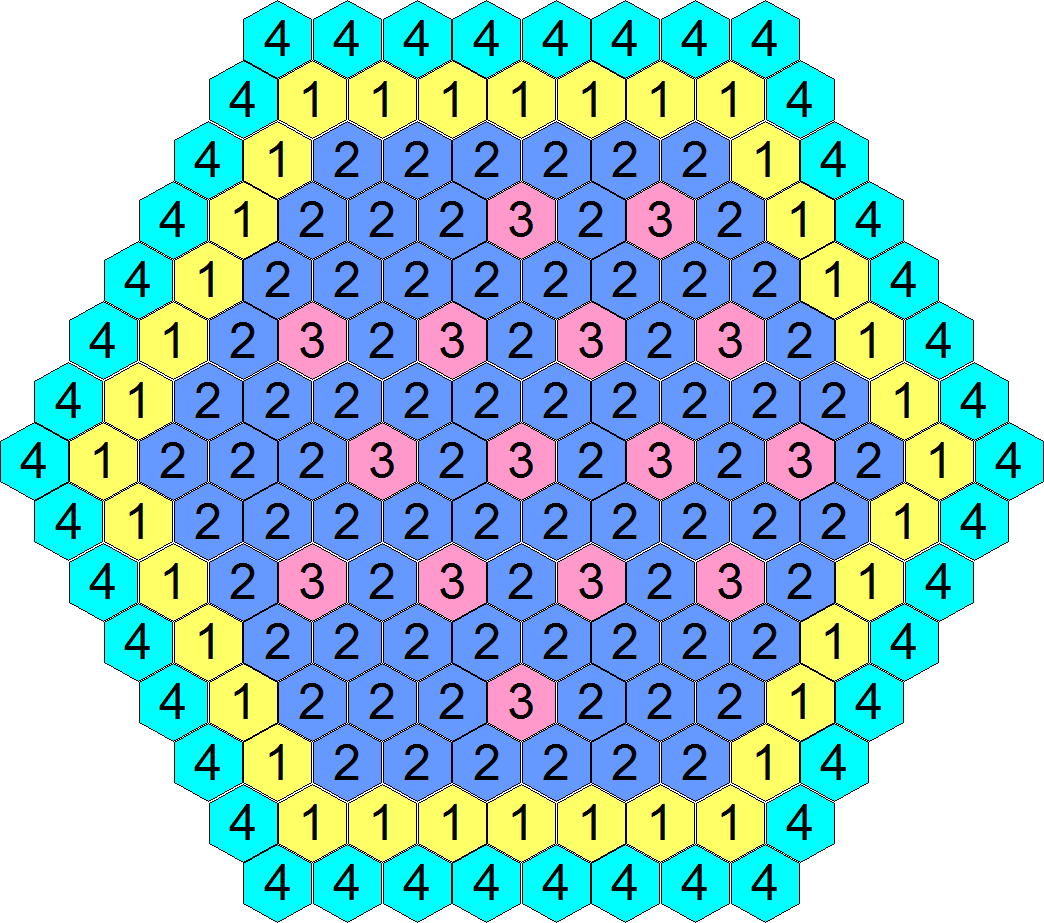}\\
		\caption{Geometrcial model of the non-symmetric test IAEA-2D.}
		\label{fig:iaea_cosym}
	\end{center}
\end{figure}

To investigate azimutally non-symmetric geometry effects on the eigenfunction behaviour, we replaced two unrodded assemblies in the north-east part of the core by rodded ones (material 3, see Fig. \ref{fig:iaea_cosym}). 

\subsubsection{Solution of Lambda Modes spectral problem}

As a reference solution for both the diffusion and transport $\mathrm{SP_3}$ models, we used the solutions obtained using very fine mesh ($p = 3, \  n = 96$). 
The effective multiplication factors are shown in Table~\ref{tab:iaea_cosym_lambda}. 
The results of the first 10 eigenvalues for $ p = 3, \  n = 96 $ are presented in Table~\ref{tab:iaea_cosym_lambda_10}.
As can be see from Table~\ref{tab:iaea_cosym_lambda_10} compared with Table~\ref{tab:iaea_with_lambda_10} for unperturbed case, all eigenvalues became well separated 
(pairs of the eigenvalues are vanished).

\begin{table}[H]
\caption{The effective multiplication factor.}
\label{tab:iaea_cosym_lambda}
\begin{center}
\begin{tabular}{r r r r r r}
\hline
$n$ & $p$ & $k_{dif}$ & $\Delta_{dif}$ &$k_{sp_3}$& $\Delta_{sp_3}$ \\
\hline
	& 1	& 1.00809& 509& 1.00931& 556\\
6	& 2	& 1.00374&  74& 1.00465&  90\\
	& 3	& 1.00306&   6& 1.00387&  12\\
\hline
	& 1	& 1.00454& 154& 1.00550& 175\\
24& 2	& 1.00310&  10& 1.00391&  16\\
	& 3	& 1.00300&   0& 1.00376&   1\\ 
\hline
	& 1	& 1.00341&  41& 1.00424&  49\\
96& 2	& 1.00300&   0& 1.00377&   2\\
	& 3	& 1.00300&  --& 1.00375&  --\\ 
\hline
Ref.&   & 1.00300&    & 1.00375&    \\ 
\hline
\end{tabular}
\end{center}
\end{table}

\begin{table}[H]
\caption{The eigenvalues $k_i=1/\lambda_i^{(k)}$ for $p=3, n=96$.}
\label{tab:iaea_cosym_lambda_10}
\begin{center}
\begin{tabular}{rrr}
\hline
$i$ & diffusion & SP$_3$  \\
\hline
1 & 1.002996 + 0.0$i$ & 1.003751 + 0.0$i$\\
2 & 0.994571 + 0.0$i$ & 0.995365 + 0.0$i$\\
3 & 0.986297 + 0.0$i$ & 0.987243 + 0.0$i$\\
4 & 0.970315 + 0.0$i$ & 0.971407 + 0.0$i$\\
5 & 0.968980 + 0.0$i$ & 0.970207 + 0.0$i$\\
6 & 0.945551 + 0.0$i$ & 0.947166 + 0.0$i$\\
7 & 0.928439 + 0.0$i$ & 0.930441 + 0.0$i$\\
8 & 0.923863 + 0.0$i$ & 0.925611 + 0.0$i$\\
9 & 0.903265 + 0.0$i$ & 0.905868 + 0.0$i$\\
10 & 0.901593 + 0.0$i$ & 0.904253 + 0.0$i$\\
\hline
\end{tabular}
\end{center}
\end{table}

\subsubsection{Solution of $\alpha$-spectral problem without delayed neutrons}

As a reference solution for both the diffusion and  $\mathrm{SP_3}$ models, we used the solutions obtained using very fine mesh ($p = 3, \  n = 96$).
The $\alpha$-spectral problem results are shown in Table~\ref{tab:iaea_cosym_alpha}.
The results of the first 10 eigenvalues for $ p = 3, \  n = 96 $ are presented in Table~\ref{tab:iaea_cosym_alpha_10}.

\begin{table}[h]
\caption{The $\alpha$-eigenvalues.}
\label{tab:iaea_cosym_alpha}
\begin{center}
\begin{tabular}{rrrrrr}
\hline
$n$ & $p$ & $\alpha_{dif}$ & $\Delta_{dif}$ &$\alpha_{sp_3}$& $\Delta_{sp_3}$ \\
\hline
	& 1	&$-$143.12 &  88.55 & $-$164.63& 96.08\\
6	& 2	& $-$67.99 &  13.42 & $-$84.75 & 16.20\\
	& 3	& $-$55.82 &   1.25 & $-$70.78 &  2.23\\ 
\hline
	& 1	& $-$81.93 &  27.36 & $-$99.47 & 30.92\\
24& 2	& $-$56.45 &   1.88 & $-$71.41 & 2.86\\
	& 3	& $-$54.65 &   0.08 & $-$68.80 & 0.25\\ 
\hline
	& 1	& $-$62.00 &   7.43 & $-$77.28 & 8.73\\
96& 2	& $-$54.74 &   0.17 & $-$68.91 & 0.36\\
	& 3	& $-$54.57 &	   -- & $-$68.55 & -- \\ 
\hline
Ref.& & $-$54.57 & & $-$68.55 \\ 
\hline
\end{tabular}
\end{center}
\end{table}

\begin{table}[h]
\caption{The eigenvalues $\alpha_i=\lambda_i^{(\alpha)}$ for $p=3, n=96$.}
\label{tab:iaea_cosym_alpha_10}
\begin{center}
\begin{tabular}{crr}
\hline
$i$ & Diffusion & SP$_3$ \\
\hline
1 &-54.57 + 0.0$i$&-68.55 + 0.0$i$ \\
2 &  97.07 + 0.0$i$&83.18 + 0.0$i$ \\
3 &242.22 + 0.0$i$&226.42 + 0.0$i$ \\
4 &513.07 + 0.0$i$&496.61 + 0.0$i$ \\
5 &530.98 + 0.0$i$&512.74 + 0.0$i$ \\
6 &898.88 + 0.0$i$&878.37 + 0.0$i$ \\
7 &1148.46 + 0.0$i$&1125.66 + 0.0$i$ \\
8 &1481.13 + 0.0$i$&1449.58 + 0.0$i$ \\
9 &1512.16 + 0.0$i$&1486.05 + 0.0$i$ \\
10&1527.83 + 0.0$i$&1501.83 + 0.0$i$ \\
\hline
\end{tabular}
\end{center}
\end{table}

\begin{figure}[H]
\begin{center}
	\includegraphics[width=0.49\linewidth]{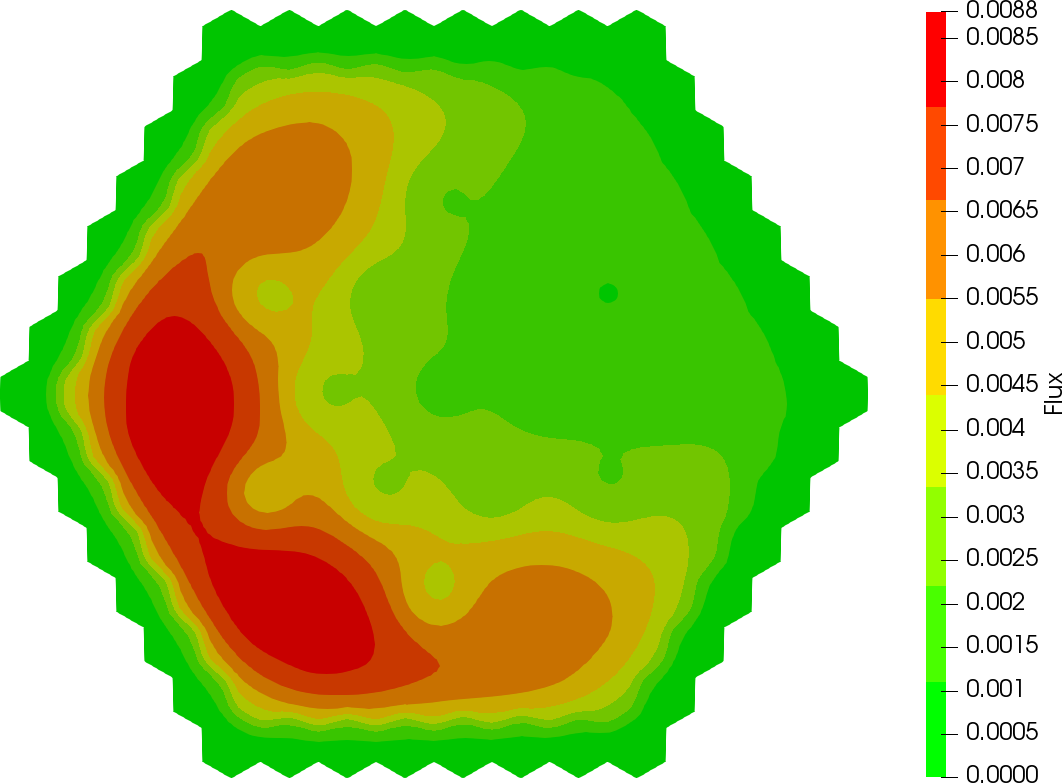}
	\includegraphics[width=0.49\linewidth]{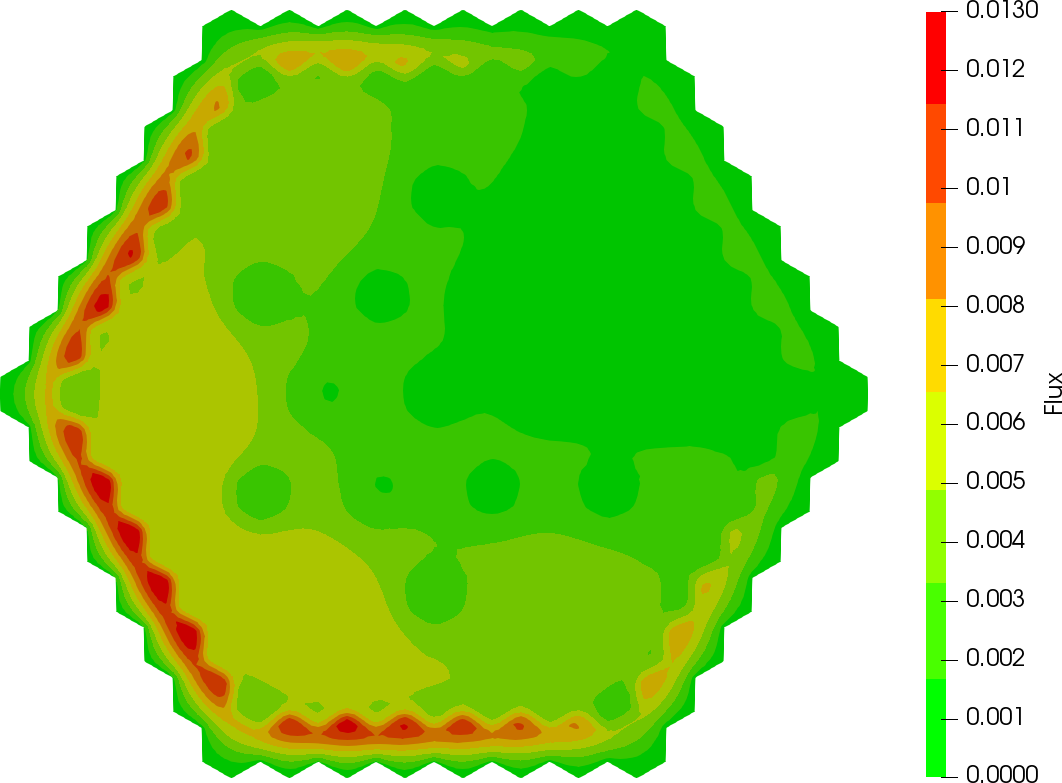}\\
	\caption{Eigenfunctions $\phi_1^{(1)}$, $\phi_2^{(1)}$.}
	\label{fig:iaea_cosym_fun_1}
\end{center}
\end{figure}

\begin{figure}[H]
\begin{center}
	\includegraphics[width=0.49\linewidth]{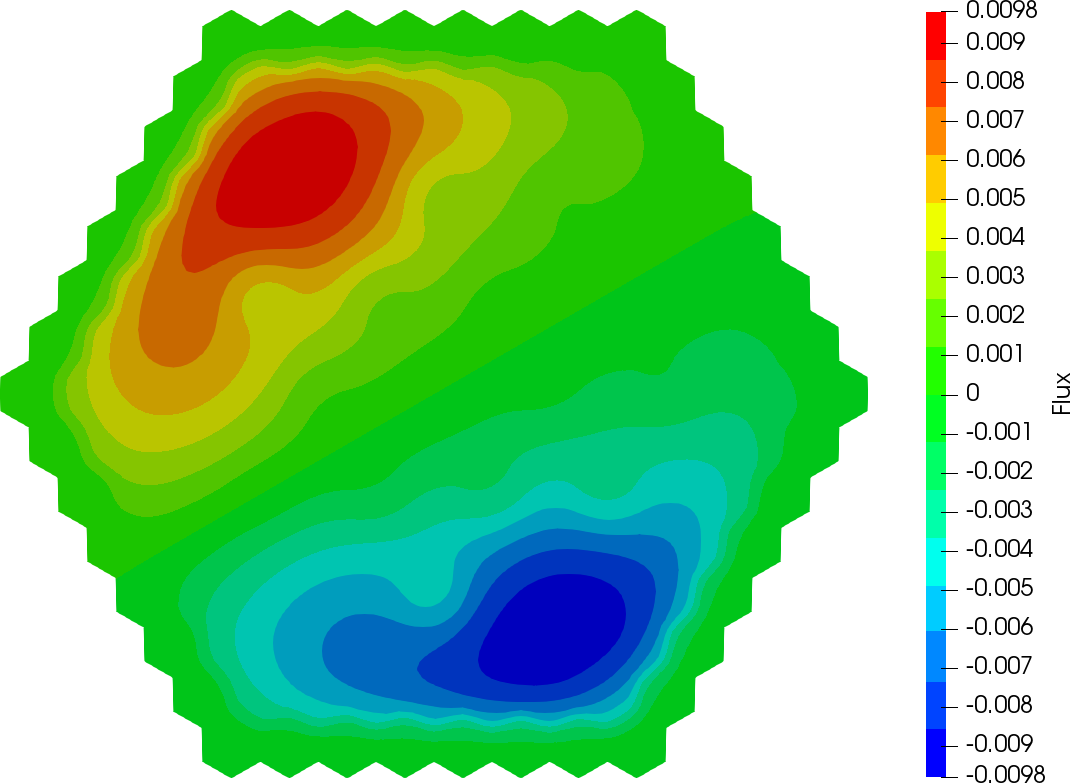}
	\includegraphics[width=0.49\linewidth]{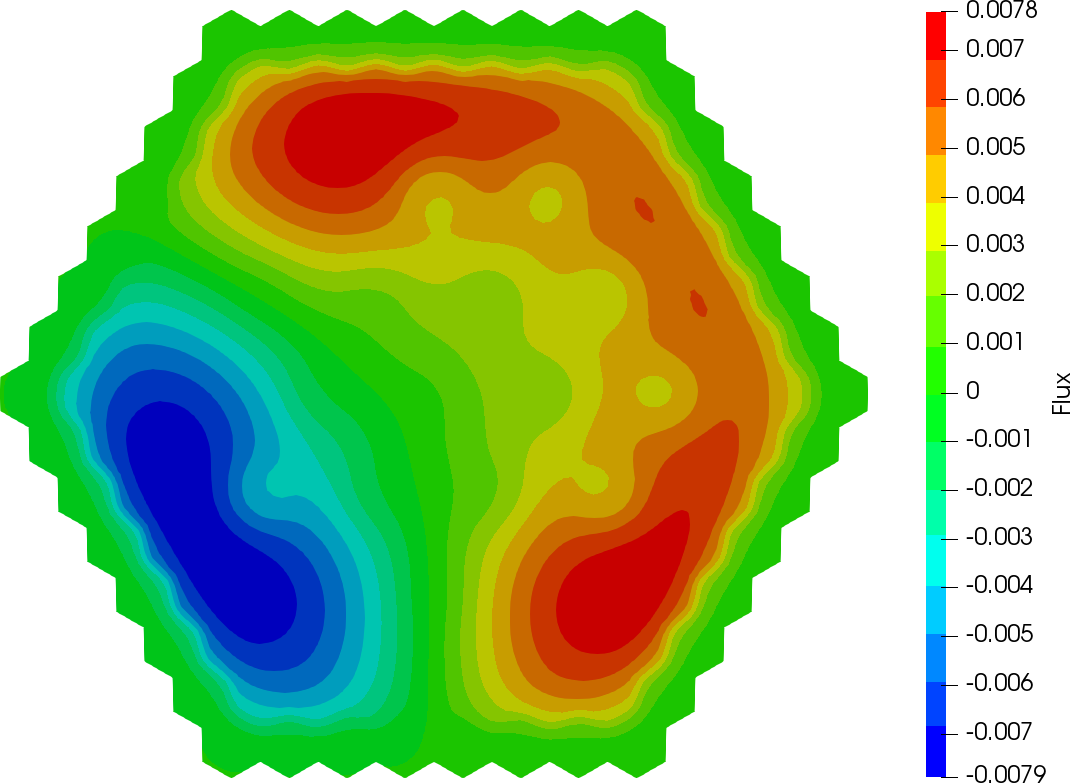}\\
	\caption{Eigenfunctions $\phi_1^{(2)}$, $\phi_1^{(3)}$.}
	\label{fig:iaea_cosym_fun_2}
\end{center}
\end{figure}

\begin{figure}[H]
\begin{center}
	\includegraphics[width=0.49\linewidth]{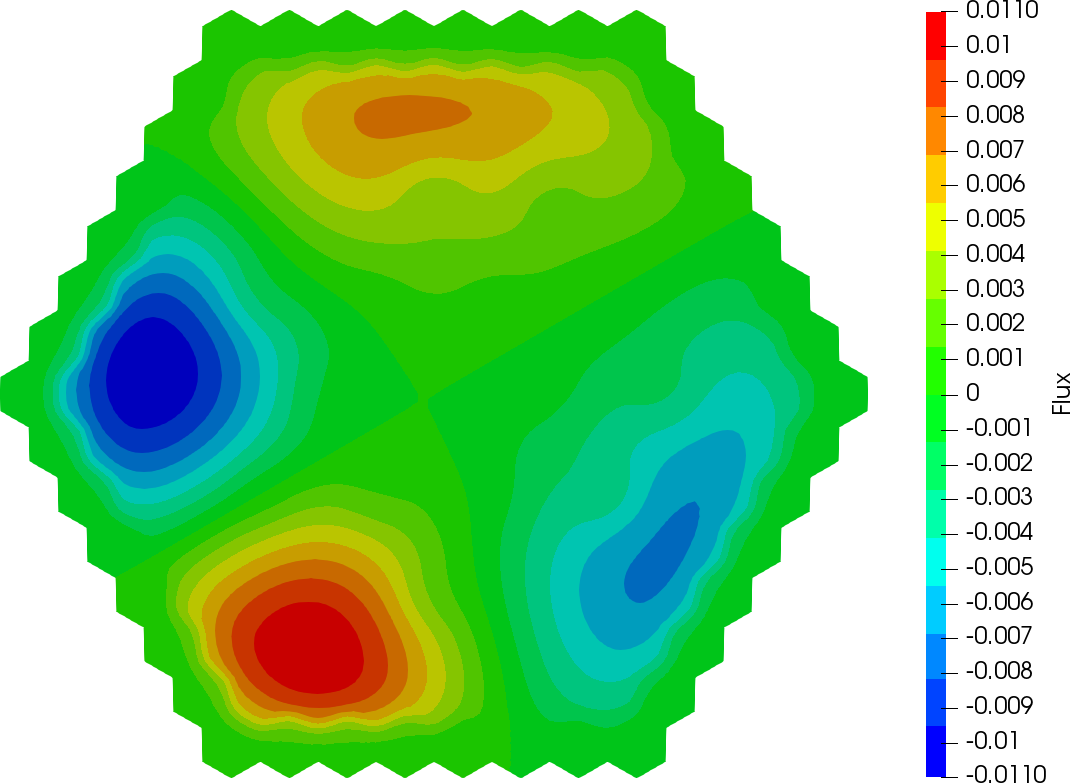}
	\includegraphics[width=0.49\linewidth]{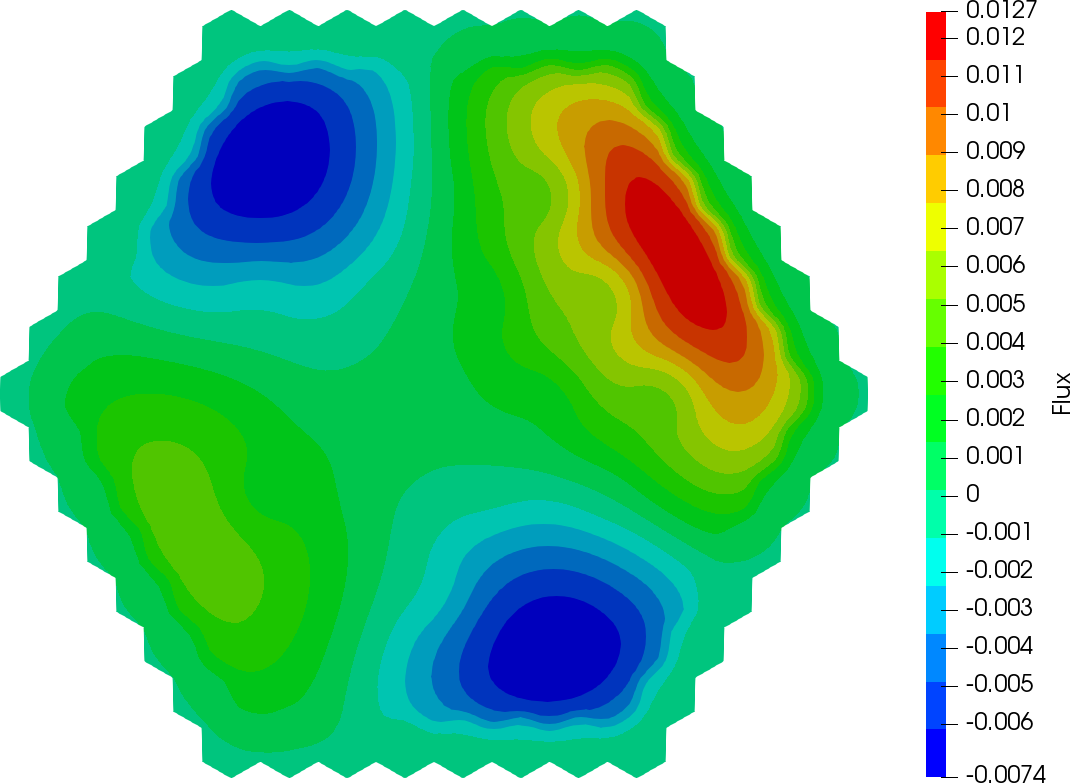}\\
	\caption{Eigenfunctions $\phi_1^{(4)}$, $\phi_1^{(5)}$.}
	\label{fig:iaea_cosym_fun_3}
\end{center}
\end{figure}

Let's consider changes in the eigenfunctions due to the rodded assembly insertion.
The eigenfunctions for fundamental eigenvalue ($i=1$) of the $\alpha$-spectral problem without delayed neutrons are shown in Fig.~\ref{fig:iaea_cosym_fun_1}. 
The eigenfunctions $\phi_1^{(i)}, i=2,3,4,5$ are shown in  Fig.~\ref{fig:iaea_cosym_fun_2}, Fig.~\ref{fig:iaea_cosym_fun_3}.
As can be seen from Fig.~\ref{fig:iaea_cosym_fun_1} to Fig.~\ref{fig:iaea_cosym_fun_3}, the overal structure of the eigenfunctions is preserved taking into account the neutron flux perturbations.


According to (\ref{1.11})-(\ref{1.13}), the prompt neutron generation time $\Lambda_{pr} = 5.474 \cdot 10^{-5} s$  for  the diffusion fundamental eigenvalue and $\Lambda_{pr} = 5.451 \cdot 10^{-5} s$  for the $\mathrm{SP_3}$ fundamental eigenvalue. Thus, for the non-symmetric test we obtained similar neutronic properties compared with the symmetric test.
The corresponding fundamental eigenvalues of the $\alpha$-spectral problem with delayed neutrons are calculated using (\ref{1.13}):
$\alpha_{dif} = -0.0679$ and
$\alpha_{sp_3} = -0.1078$.

\subsection{HWR test problem}

\begin{figure}[h]
	\begin{center}
    		\includegraphics[width=0.9\linewidth] {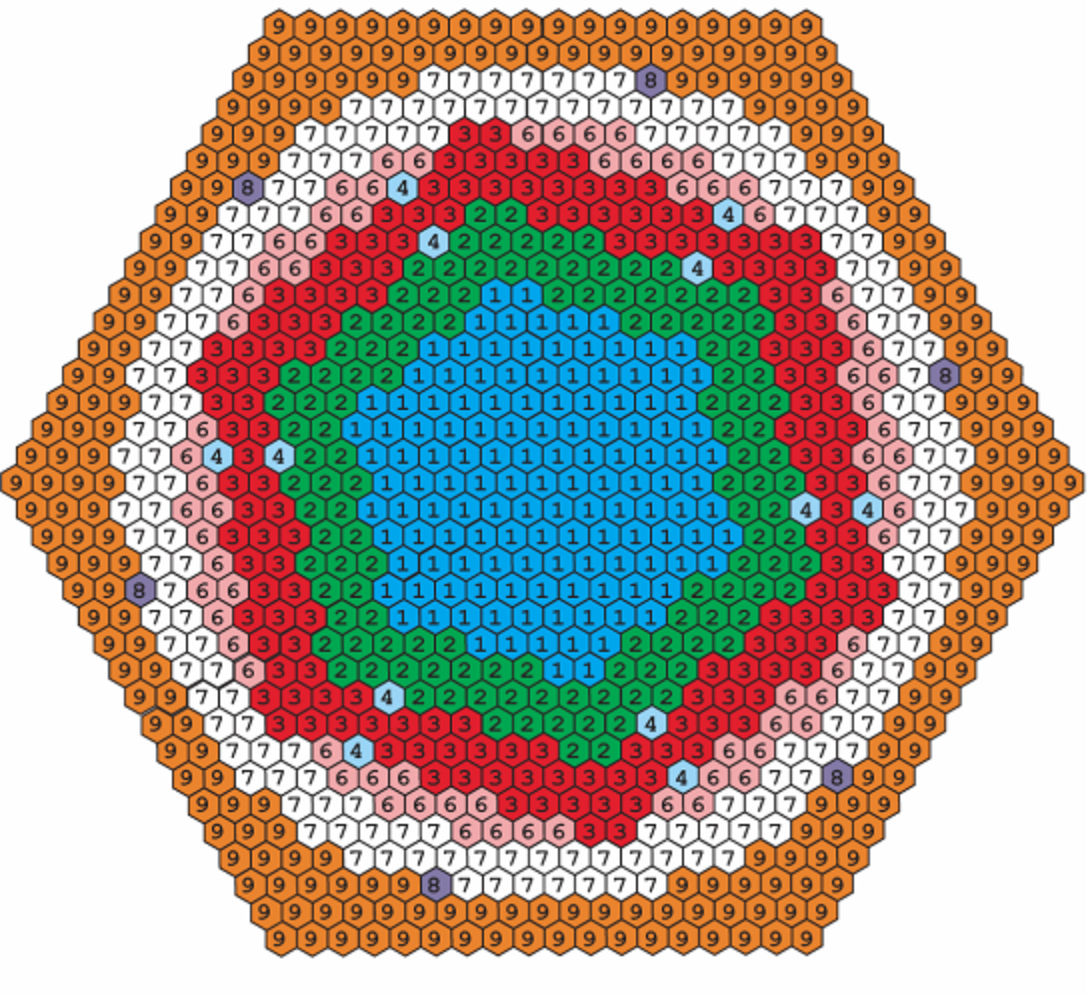}
		\caption{Geometrcial model of the HWR test.}
		\label{fig:hwr}
	\end{center}
\end{figure}

This benchmark is a model of large heavy-water reactor HWR \citep{chao1995}. 
The geometry of the HWR test is presented in Fig. \ref{fig:hwr}. Fuel assemblis (1, 2, 3 and 6 in Fig. \ref{fig:hwr}) located in the central part of the core, are surrouded by the target zone and reflector layer (7 and 9 in Fig. \ref{fig:hwr}). There are two types of rodded assemblies (4 and 8).
The assembly size is equal to 17.78 cm. 

Diffusion constants  are given in Table \ref{tab:hwr}. 
The following delayed neutrons parameters are used: one group of delayed neutrons with effective fraction $\beta_1 = 6.5\cdot10^{-3}$ and decay constant $\lambda_1 = 0.08$ s$^{-1}$. 
Neutron velocity  $v_1 = 1.25 \cdot 10^7$ cm/s and $v_2 = 2.5 \cdot 10^5$ cm/s.

\begin{table}[h]
\caption{Diffusion constants for HWR test.}
\label{tab:hwr}
\begin{center}
\begin{tabular}{ccllll}
\hline
Material & Group & $D$, cm & $\Sigma_r$, cm$^{-1}$ & $\Sigma_{1\to 2}$, cm$^{-1}$ & $\nu\Sigma_f$, cm$^{-1}$\\
\hline
\multirow{ 2}{*}{1} & 1 & 1.38250058 & 1.1105805e-2 & \multirow{ 2}{*}{8.16457e-3} & 2.26216e-3 \\
  & 2 & 0.89752185 & 2.2306487e-2 &            & 2.30623e-2 \\
\hline
\multirow{ 2}{*}{2} & 1 & 1.38255219 & 1.1174585e-2 & \multirow{ 2}{*}{8.22378e-3} & 2.22750e-3 \\
  & 2 & 0.89749043 & 2.2387609e-2 &            & 2.26849e-2 \\
\hline
\multirow{ 2}{*}{3} & 1 & 1.37441741 & 1.0620368e-2 & \multirow{ 2}{*}{8.08816e-3} & 2.14281e-3 \\
  & 2 & 0.88836771 & 1.6946527e-2 &            & 2.04887e-2 \\
\hline
\multirow{ 2}{*}{4} & 1 & 1.31197955 & 1.2687953e-2 & \multirow{ 2}{*}{1.23115e-2} & 0.0 \\
  & 2 & 0.87991376 & 5.2900925e-2 &            & 0.0 \\
\hline
\multirow{ 2}{*}{6} & 1 & 1.38138909 & 1.056312e-2 & \multirow{ 2}{*}{7.76568e-3} & 2.39469e-3 \\
  & 2 & 0.90367052 & 2.190298e-2 &            & 2.66211e-2 \\
\hline
\multirow{ 2}{*}{7} & 1 & 1.30599110 & 1.1731321e-2 & \multirow{ 2}{*}{1.10975e-2} & 0.0 \\
  & 2 & 0.83725587 & 4.3330365e-3 &            & 0.0 \\
\hline
\multirow{ 2}{*}{8} & 1 & 1.29192957 & 1.1915316e-2 & \multirow{ 2}{*}{1.15582e-2} & 0.0 \\
  & 2 & 0.81934103 & 3.0056488e-4 &            & 0.0 \\
\hline
\multirow{ 2}{*}{9} & 1 & 1.06509884 & 2.8346221e-2 & \multirow{ 2}{*}{2.61980e-2} & 0.0 \\
  & 2 & 0.32282849 & 3.3348874e-2 &            & 0.0 \\  
\hline
\end{tabular}
\end{center}
\end{table}

\subsubsection{Solution of Lambda Modes spectral problem}
\begin{table}[H]
\caption{The effective multiplication factor.}
\label{tab:hwr_lambda}
\begin{center}
\begin{tabular}{c c r r r r r r}
\hline
$n$ & $p$ & $k_{dif}$ & $\Delta_{dif},pcm$ & $\delta_{dif}$ &$k_{sp_3}$& $\Delta_{sp_3},pcm$ & $\delta_{dif}$ \\
\hline
	& 1	& 0.991985&  2.0& 1.16&0.992178&   5.0& 0.80\\
6	& 2	& 0.991989&  2.4& 0.31&0.992166&   3.8& 0.24\\
	& 3	& 0.991964&  0.1& 0.08&0.992132&   0.4& 0.07\\
\hline
	& 1	& 0.991983&  1.8& 0.05&0.992165&   3.7& 0.08\\
24& 2	& 0.991965&  0.0& 0.01&0.992133&   0.5& 0.01\\
	& 3	& 0.991963&  0.2& 0.01&0.992128&   0.0& 0.00\\ 
\hline
	& 1	& 0.991969&  0.4& 0.08&0.992140&   1.2& 0.01\\
96& 2	& 0.991963&  0.2& 0.02&0.992129&   0.1& 0.00\\
	& 3	& 0.991963&  0.2& 0.01&0.992128&    --& --\\ 
\hline
Ref.&   & 0.991965&    & & 0.992128 & & \\ 
\hline
\end{tabular}
\end{center}
\end{table}

As a reference solution for the diffusion model, we used the  results obtained by \citep{chao1995}; for the $\mathrm{SP_3}$ model --- the solution obtained using very fine mesh ($p = 3, \  n = 96$).

The effective multiplication factors for the HWR test are shown in Table~\ref{tab:hwr_lambda}. 

\begin{table}[h]
\caption{The eigenvalues $k_i=1/\lambda_i^{(k)}$ for $p=3, n=96$.}
\label{tab:hwr_lambda_10}
\begin{center}
\begin{tabular}{c l l }
\hline
$i$ & diffusion & SP$_3$  \\
\hline
1 & 0.991963 + 0.0$i$   & 0.992128 + 0.0$i$\\
2 & 0.983594 + 1.1645e-05$i$   & 0.983793 + 1.2072e-05$i$\\
3 & 0.983594 $-$ 1.1645e-05$i$ & 0.983793 $-$ 1.2072e-05$i$\\
4 & 0.964240 + 2.1564e-05$i$   & 0.964523 + 2.2337e-05$i$\\
5 & 0.964240 $-$ 2.1564e-05$i$ & 0.964523 $-$ 2.2337e-05$i$\\
6 & 0.943290 + 0.0$i$   & 0.943733 + 0.0$i$\\
7 & 0.923872 + 0.0$i$   & 0.924257 + 0.0$i$\\
8 & 0.918657 + 0.0$i$   & 0.918798 + 0.0$i$\\
9 & 0.895682 + 3.5570e-05$i$   & 0.896317 + 3.6750e-05$i$\\
10 & 0.895682 $-$ 3.5570e-05$i$& 0.896317 + 3.6750e-05$i$\\
\hline
\end{tabular}
\end{center}
\end{table}

The results of the first 10 eigenvalues for $ p = 3, \  n = 96 $ are presented in Table~\ref{tab:hwr_lambda_10}.
The eigenvalues $k_2, k_3, k_4, k_5, k_9, k_{10}$ of the $\lambda$-spectral problem are the complex values with small imaginary parts, and the eigenvalues $k_1, k_6, k_7, k_8$ are the real values. One can see pairs of the complex
conjugate values.

\subsubsection{Solution of $\alpha$-spectral problem without delayed neutrons}

As a reference solution for the diffusion and $\mathrm{SP_3}$ models we use the solutions obtained using very fine mesh ($p = 3, \  n = 96$).

The $\alpha$-spectral problem results at different computational parameters are shown in Table~\ref{tab:hwr_alpha}. 
The results of the first 10 eigenvalues for $p = 3, \  n = 96 $ are presented in Table~\ref{tab:hwr_alpha_10}.
The eigenvalues are well separated.
The eigenvalues $\alpha_2, \alpha_3$, $\alpha_4, \alpha_5$, $\alpha_9, \alpha_{10}$ of the $\alpha$-spectral problem, like for the $\lambda$-spectral problem, are the complex values with small imaginary parts, and the eigenvalues $\alpha_1, \alpha_6$, $\alpha_7, \alpha_8$ are the real values.

\begin{figure}[H]
\begin{center}
	\includegraphics[width=0.49\linewidth]{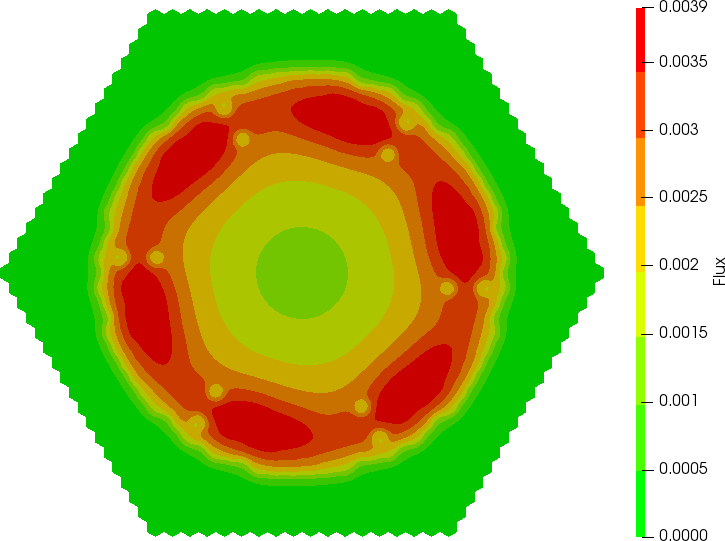}
	\includegraphics[width=0.49\linewidth]{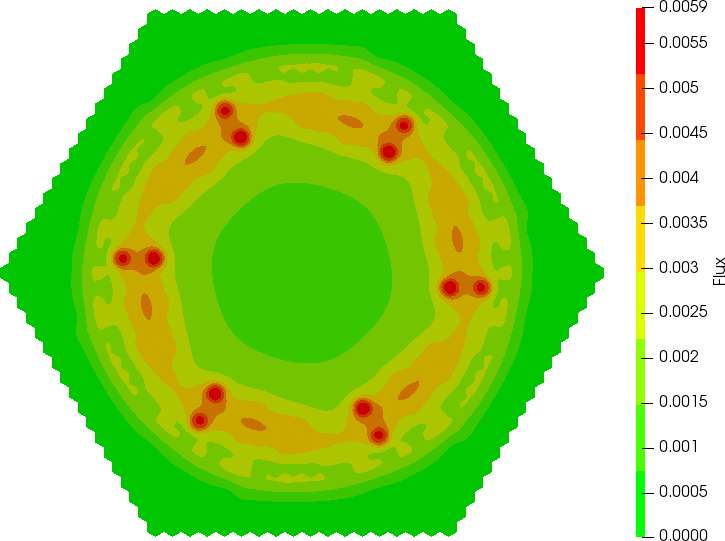}\\
	\caption{The eigenfunctions $\phi^{(1)}_1$ (left) and $\phi^{(1)}_2$ (right).}
	\label{fig:hwr_fun_1}
\end{center}
\end{figure}

\begin{figure}[H]
\begin{center}
	\includegraphics[width=0.49\linewidth]{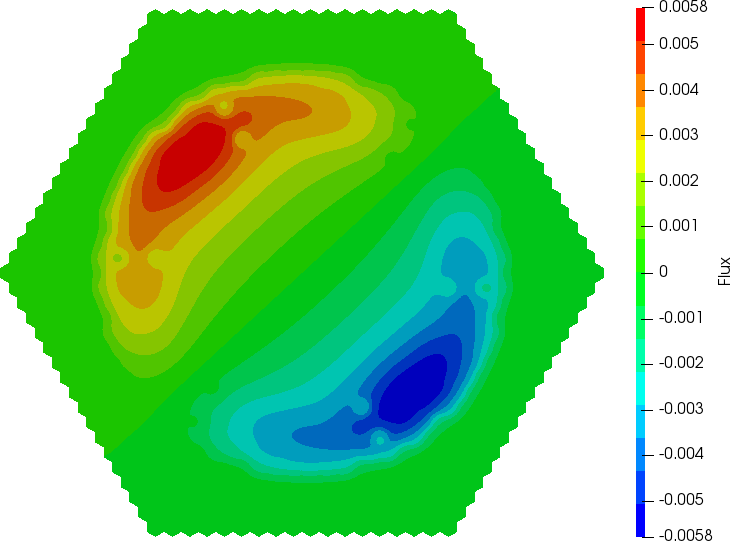}
	\includegraphics[width=0.49\linewidth]{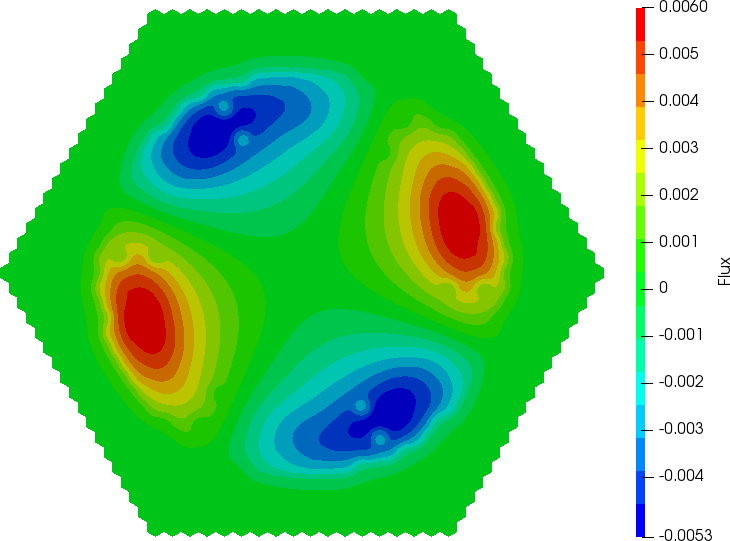}\\
	\caption{Real part of eigenfunctions $\phi^{(2)}_1, \ \phi^{(3)}_1$ (left) and $\phi^{(4)}_1, \ \phi^{(5)}_1$ (right).}
	\label{fig:hwr_fun_2}
\end{center}
\end{figure}

\begin{figure}[H]
\begin{center}
	\includegraphics[width=0.49\linewidth]{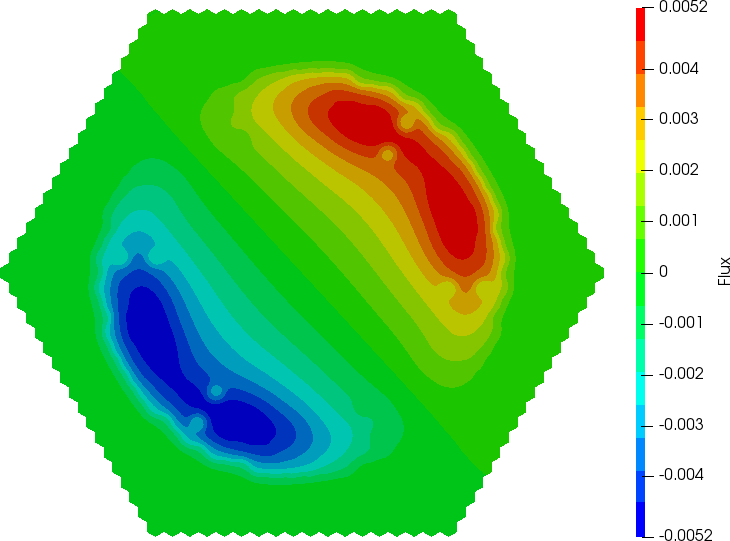}
	\includegraphics[width=0.49\linewidth]{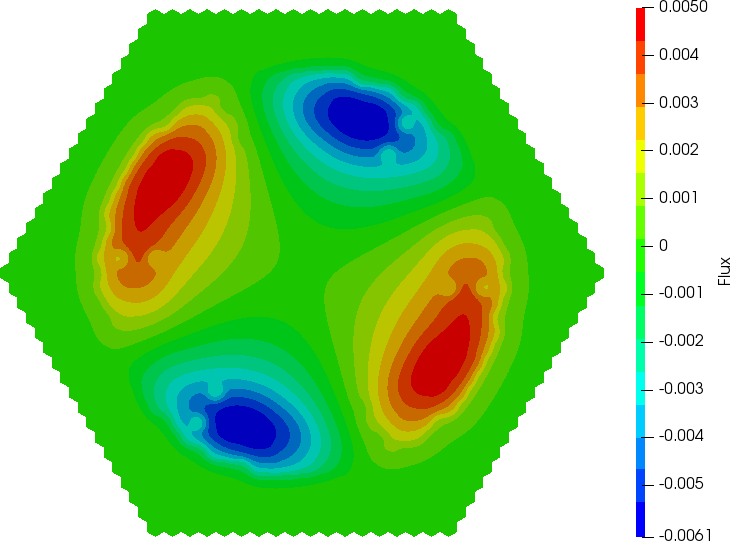}\\
	\caption{Imaginary part of eigenfunctions $\phi^{(2)}_1, \ - \phi^{(3)}_1$ (left) and $\phi^{(4)}_1, \ - \phi^{(5)}_1$ (right).}
	\label{fig:hwr_fun_3}
\end{center}
\end{figure}

The eigenfunctions for fundamental eigenvalue ($n=1$) of the $\alpha$-spectral problem  are shown in Fig.~\ref{fig:hwr_fun_1}.
The real part of the eigenfunctions $\phi^{(n)}_1, \ n = 2,3,4,5$ is shown in Fig.~\ref{fig:hwr_fun_2}.
Fig.~\ref{fig:hwr_fun_3} shows the imaginary part of these eigenfunctions.
The eigenfunctions of the $\lambda$-spectral and $\alpha$-spectral problems are close to each other in topology.

\begin{table}[h]
\caption{The $\alpha$-eigenvalues.}
\label{tab:hwr_alpha}
\begin{center}
\begin{tabular}{rrrrrr}
\hline
$n$ & $p$ & $\alpha_{dif}$ & $\Delta_{dif}$ &$\alpha_{sp_3}$& $\Delta_{sp_3}$ \\
\hline
	& 1	&42.281 & 0.018 & 41.246 & 0.134\\
6	& 2	&42.135 & 0.128 & 41.190 & 0.190\\
	& 3	&42.259 & 0.004 & 41.362 & 0.018\\ 
\hline
	& 1	&42.196 & 0.067 & 41.228 & 0.152\\
24& 2	&42.253 & 0.010 & 41.354 & 0.026\\
	& 3	&42.263 & 0.000 & 41.379 & 0.001\\ 
\hline
	& 1	&42.241 & 0.022 & 41.330 & 0.050\\
96& 2	&42.262 & 0.001 & 41.377 & 0.003\\
	& 3	&42.263 &    -- & 41.380 & -- \\ 
\hline
Ref.& & 42.263 & & 41.380 \\ 
\hline
\end{tabular}
\end{center}
\end{table}

\begin{table}[h]
\caption{The eigenvalues $\alpha_i=\lambda_i^{(\alpha)}$ for $p=3, n=96$.}
\label{tab:hwr_alpha_10}
\begin{center}
\begin{tabular}{c l l}
\hline
$i$ & Diffusion & SP$_3$ \\
\hline
1 &\phantom{0}42.263 + 0.0$i$       &41.380 + 0.0$i$ \\
2 &\phantom{0}84.867 $-$ 0.06130$i$ &83.821 $-$ 0.06358$i$ \\
3 &\phantom{0}84.867 + 0.06130$i$   &83.821 + 0.06358$i$ \\
4 &182.914 $-$ 0.11367$i$           &181.471 $-$ 0.11805$i$ \\
5 &182.914 + 0.11367$i$             &181.471 + 0.11805$i$ \\
6 &293.017 + 0.0$i$                 &290.940 + 0.0$i$ \\
7 &371.528 + 0.0$i$                 &369.374 + 0.0$i$ \\
8 &515.465 $-$ 0.16397$i$           &512.337 $-$ 0.17197$i$ \\
9 &515.465 + 0.16397$i$             &512.337 + 0.17197$i$ \\
10&518.670 + 0.0$i$                 &517.975 + 0.0$i$ \\
\hline
\end{tabular}
\end{center}
\end{table}


\subsubsection{Solution of $\alpha$-spectral problem with delayed neutrons}

As a reference solution for the diffusion and $\mathrm{SP_3}$ models we use the solutions obtained using very fine mesh ($p = 3, \  n = 96$).

The $\alpha$-spectral problem results are shown in Table~\ref{tab:hwr_alpha_del}. 
Due to the contribution of delayed neutrons, the fundamental eigenvalue is much smaller than in the case without delayed neutrons.
The results of the first 10 eigenvalues for $p = 3, \  n = 96 $ is shown in Table~\ref{tab:hwr_alpha_del_10}.
The eigenvalues $\alpha_2, \alpha_3$, $\alpha_4, \alpha_5$, $\alpha_9, \alpha_{10}$ of the $\alpha$-spectral problem, like as before, are the complex values with small imaginary parts, and the eigenvalues $\alpha_1, \alpha_6$, $\alpha_7, \alpha_8$ are the real values.

\begin{table}[h]
\caption{The $\alpha$-eigenvalues.}
\label{tab:hwr_alpha_del}
\begin{center}
\begin{tabular}{rrrrrr}
\hline
$n$ & $p$ & $\alpha_{dif}$ & $\Delta_{dif}$ &$\alpha_{sp_3}$& $\Delta_{sp_3}$ \\
\hline
	& 1	&0.04431 & 0.00006 & 0.04383 & 0.00012\\
6	& 2	&0.04430 & 0.00007 & 0.04386 & 0.00009\\
	& 3	&0.04437 & 0.00000 & 0.04394 & 0.00001\\ 
\hline
	& 1	&0.04432 & 0.00005 & 0.04386 & 0.00009\\
24& 2	&0.04436 & 0.00001 & 0.04394 & 0.00001\\
	& 3	&0.04437 & 0.00000 & 0.04395 & 0.00000\\ 
\hline
	& 1	&0.04435 & 0.00002 & 0.04392 & 0.00003\\
96& 2	&0.04437 & 0.00000 & 0.04395 & 0.00000\\
	& 3	&0.04437 & --      & 0.04395 & -- \\ 
\hline
Ref.& & 0.04437 & & 0.04395 \\ 
\hline
\end{tabular}
\end{center}
\end{table}

\begin{table}[h]
\caption{The eigenvalues $\alpha_i=\lambda_i^{(\alpha)}$ for $p=3, n=96$.}
\label{tab:hwr_alpha_del_10}
\begin{center}
\begin{tabular}{c l l}
\hline
$i$ & Diffusion & SP$_3$ \\
\hline
1 &0.04437 + 0.0$i$     		&0.04395 + 0.0$i$ \\
2 &0.05755 $-$ 1.15549e-05$i$ 	&0.05735 $-$ 1.22333e-05$i$ \\
3 &0.05755 + 1.15549e-05$i$   	&0.05735 + 1.22333e-05$i$ \\
4 &0.06807 $-$ 6.35264e-06$i$   &0.06798 $-$ 6.66947e-06$i$ \\
5 &0.06807 + 6.35264e-06$i$     &0.06798 + 6.66947e-06$i$ \\
6 &0.07219 + 0.0$i$             &0.07213 + 0.0$i$ \\
7 &0.07415 + 0.0$i$             &0.07412 + 0.0$i$ \\
8 &0.07453 + 0.0$i$          	&0.07452 + 0.0$i$ \\
9 &0.07577 $-$ 1.52484e-06$i$   &0.07574 $-$ 1.60360e-06$i$ \\
10&0.07577 + 1.52484e-06$i$     &0.07574 + 1.60360e-06$i$ \\
\hline
\end{tabular}
\end{center}
\end{table}

According to (\ref{1.11})-(\ref{1.13}), the prompt neutron generation time $\Lambda_{pr} = 1.917 \cdot 10^{-4} s$  for both the diffusion and $\mathrm{SP_3}$ fundamental eigenvalues.
Assuming the same value of $\Lambda_{pr}$ for both $ \alpha $-spectral problems, the uncertainty in the fundamental eigenvalues is less than 0.1 percent.

\section*{Conclusion}

Simulation of reactor dynamic processes is basically considered on the basis of multigroup diffusion approximation of the neutron transport equation. 
To improve the calculation accuracy for some situations of interest, including pin-by-pin calculations, the $\mathrm{SP_3}$ approximation was adopted in different whole-core diffusion codes as an improved option compared with the diffusion approximation. 
In this regard, it will be very useful to compare the spectral parameters, calculated by both the diffusion and $\mathrm{SP_3}$ options using the finite element method. 

Solution of the $k$-spectral ($\lambda$-spectral)  and $\alpha$-spectral problems has been tested for the IAEA-2D and HWR reactor benchmark tests. 
The classical Lagrange finite elements are used for the spatial approximation. 
Accuracy control is performed using condensed grids. 
Spectral problems are solved numerically using well-developed free software SLEPc and using GMSH as a generic mesh generator.
 
Of particular interest is the problem associated with appearance of complex eigenvalues and eigenfunctions. 
It was found that this tendency occurs for both the diffusion and $\mathrm{SP_3}$ solutions of the HWR reactor test. 
On the contrary, for the IAEA-2D benchmark tests we have not  found complex eigenvalues and eigenfunctions, even for the non-symmetric variant of the benchmark.

There were found some features in the $k$-spectral and $\alpha$-eigenvalues, which were confirmed by calculations.
Using the known fundamental $k-$eigenvalue and the evaluated value of the prompt neutron generation time $\Lambda_{pr}$ one can estimate any fundamental $\alpha-$eigenvalue with and without delayed neutrons. The same behaviour occurs for the minor eigenvalues.

\section*{Acknowledgements}

This work was supported by the grant of the Russian Federation Government (\#~14.Y26.31.0013) and by the Russian Foundation for Basic Research (\#~18-31-00315).


\end{document}